\documentclass[12pt,useAMS]{article}
\usepackage{graphicx}
\usepackage{multirow}
\usepackage{enumerate}
\usepackage{varwidth}
\usepackage{amsmath, amsthm, amssymb}

\def\aln{\begin{align*}}
\def\eln{\end{align*}}
\def\be{\begin{equation}}
\def\ee{\end{equation}}
\def\bse{\begin{eqnarray*}}
\def\ese{\end{eqnarray*}}
\def\bea{\begin{eqnarray}}
\def\eea{\end{eqnarray}}

\def\var{\hbox{var}}
\def\var{\hbox{Var}}

\def\sgn{\hbox{sgn}}
\def\arg{\hbox{arg}}

\def\bxi{\boldsymbol\xi}
\def\bal{\boldsymbol\alpha}
\def\bka{\boldsymbol\kappa}
\def\bel{\boldsymbol\epsilon}
\def\bbe{\boldsymbol\beta}
\def\bdx{\boldsymbol x}
\def\mbx{\mathbf x}
\def\mby{\mathbf y}
\def\ma{\mathcal{A}}
\def\refmark{\par\vskip 2mm\noindent\refhg}
\def\refhg{\hangindent=20pt\hangafter=1}

\def\JASA{{\it Journal of the American Statistical Association}}

\def\ANNALS{{\it The Annals of Statistics}}

\def\STATSCI{{\it Statistical Science}}

\def\JRSSB{{\it Journal of the Royal Statistical Society, Series B}}
\def\JMLR{{\it Journal of Machine Learning Research}}
\def\BMCS{{\it Biometrics}}

\def\TECH{{\it Technometrics}}

\def\SSS{{\it Statistica Sinica}}

\newcommand{\Appendix}{\appendix\def\thesection{Appendix~\Alph{section}}\def\thesubsection{\Alph{section}.\arabic{subsection}}}
\begin{document}

\font\fmain=cmbx10 scaled\magstep4
\title{Flexible Variable Selection for Recovering Sparsity in Nonadditive Nonparametric Models}
\author{Zaili Fang$^1$, Inyoung Kim$^{1*}$, and Patrick Schaumont$^2$}
\date{\today}
\maketitle
\thispagestyle{empty}
\baselineskip=26pt
\hskip 5mm \\
1. Department of Statistics, Virginia Polytechnic
  Institute and State University, Blacksburg, Virginia, U.S.A.\\
2. Department of Electrical \& Computer Engineering, Virginia Polytechnic
  Institute and State University, Blacksburg, Virginia, U.S.A.\\
\hskip 5mm \\
\hskip 5mm \\
\noindent
*To whom correspondence should be addressed:\\
Inyoung Kim, Ph.D.\\
Department of Statistics, Virginia Polytechnic
  Institute and State University, 410A Hutcheson Hall, Blacksburg, VA 24061-0439, U.S.A.\\
Tel: (540) 231-5366\\
Fax: (540) 231-3863\\
Email: inyoungk$@$vt.edu\\
\hskip 5mm \\

\newpage
\begin{abstract}
%\begin{center}
%{\large{\bf SUMMARY}}
%\end{center}
Variable selection for recovering sparsity in nonadditive nonparametric models has been challenging. This problem becomes even more difficult due to complications in modeling unknown interaction terms among high dimensional variables. There is currently no variable selection method to overcome these limitations. Hence, in this paper we propose a variable selection approach that is developed by connecting a kernel machine with the nonparametric multiple regression model. The advantages of our approach are that it can: (1) recover the sparsity, (2) automatically model unknown and complicated interactions, (3) connect with several existing approaches including linear nonnegative garrote, kernel learning and automatic relevant determinants (ARD), and (4) provide flexibility for both additive and nonadditive nonparametric models. Our approach may be viewed as a nonlinear version of a nonnegative garrote method. We model the smoothing function by a least squares kernel machine and construct the nonnegative garrote objective function as the function of the similarity matrix. Since the multiple regression similarity matrix can be written as an additive form of univariate similarity matrices corresponding to input variables, applying a sparse scale parameter on each univariate similarity matrix can reveal its relevance to the response variable. We also derive the asymptotic properties of our approach, and show that it provides a square root consistent estimator of the scale parameters. Furthermore, we prove that sparsistency is satisfied with consistent initial kernel function coefficients under certain conditions and give the necessary and sufficient conditions for sparsistency. An efficient coordinate descent/backfitting algorithm is developed. A resampling procedure for our variable selection methodology is also proposed to improve power.
\vskip 5mm
\noindent
\underline{\hbox{\bf Keywords}}: Automatic Relevant Determinant; Kernel machine; LASSO; Multivariate smoothing function; Nonnegative garrote; Sparsistency; Variable selection.
\vskip 5mm
\noindent
\underline{\hbox{\bf Running Title} }: Flexible Variable Selection in Multivariate Nonparametric Models
\thispagestyle{empty}
\end{abstract}
\newpage

%%%%%%%%%%%%%%%%%%%%%%%%%%%%%%%%%%%%%%%%%%%%%%%%%%%%%%%%%%%%%%%%%%%%%%%%%%%%%%%

\section{Introduction}\label{sec1}
The variable selection problem is important in many research areas such as genomics, data mining, image analysis, text and speech analysis, and other areas with high dimensional data. In general, the input variables form an interacting network with one another and modeling these interactions is complicated due to high order interaction terms.

There are numerous approaches to modeling high dimensional data using multi-dimensional nonparametric models (Wahba, 1990; Green and Silverman, 1994; Hastie and Tibshirani, 1990). Most variable selection approaches in multi-dimensional nonparametric models are performed in terms of function components selection, that is, modeling the function components (including nonlinear interactions) additively and then selecting significant components. Examples of these variable selection approaches are Component Selection and Smoothing Operator (COSSO) (Lin and Zhang, 2006), Sparse Additive Models (SAMs) (Ravikumar et al., 2009), and the extension of SAMs, Variable Selection using Adaptive Nonlinear Interaction Structure in High dimensions (VANISH) (Radchenko and James, 2010). However, when the number of input variables is large and their interactions are complicated, modeling each interaction term is extremely expensive and these function components approaches may not be efficient.

Other variable selection approaches based on the kernel machine method have achieved great success.  Liu et al. (2007) established the connection between the least squares kernel machine (LSKM) and linear mixed models. Zou et al. (2010) employed a nonparametric regression model with a Gaussian process which simultaneously considers all possible interactions. In these works, the interactions among the multi-dimensional variables are modeled automatically by the kernel. Because of their simplicity and generality, function kernels and associated function spaces are a powerful technique to analyze multi-dimensional data.

In this paper we will also focus on variable selection approaches based on the kernel machine method because the family of kernel functions is extremely rich for multiple regression smoothing. They have the flexibility for various models including additive functional ANOVA and nonadditive smoothing functions.  For example, since any symmetric positive definite matrix is a valid Gram matrix,
an additive Gram matrix $K=\sum_j^p\xi_jK_j$ ($\xi_j$'s are nonnegative hyperparameters) can be used for the functional ANOVA $f\left(\bdx^T\right)=\sum_{j=1}^pf_j(x_j)$, where $K_j$ is the Gram matrix for the $j$th function space $f_j$ and $\bdx^T=(x_1,...,x_p)$. According to the the Representer Theorem (Kimeldorf and Wahba, 1971), a nonparametric function can be represented using a kernel function, $f_j(x)=\sum_{l=1}^n\alpha_lk_j(x_l,x)$ (the dual representation), where the $\alpha_l$'s are the kernel function coefficients. With penalty on the norm (or pseudonorm) of the $j$th function component $f_j$, $\|f_j\|_{\mathcal{H}_{K_j}}$, sparsity of the function components can be recovered (Lin and Zhang, 2006; Bach, 2008). %This has been already applied to COSSO and multiple kernel learning (MKL) (Rakotomamonjy et al., 2008). Furthermore, in the MKL $K=\sum_j^p\xi_jK_j$ matrix, each $K_j$ is corresponding to a specific function $f_j$ spanned by a particular set of orthogonal basis function $\{\phi^j_l(x)\}_{l=1}^d$. That is, the function component of $j$th variable can be expressed as $f_j(x)=\sum_{l=1}^d\omega^j_l\phi^j_l(x)$ (the primal representation). Applying the penalty on $\boldsymbol{\omega}^j$ through $\sqrt{n^{-1}{\boldsymbol{\omega}^j}^T{\boldsymbol{\phi}^j}^T\boldsymbol{\phi}^j\boldsymbol{\omega}^j}$, where $\boldsymbol{\omega}^j$ is vector of $\omega^j_l$'s and $\boldsymbol{\phi}^j$ is vector of $\phi^j_l$'s, one can have the sparse additive models (SAMs) and  function components sparsity might be obtained by shrinking $f_j$ to zero. %Ravikumar et al. (2009) studied sparsistency properties of SAMs and Bach (2008) extended the consistency conditions to MKL. So far, there is only few theoretical studies on the consistency of nonnegative garrote (Yuan and Lin, 2007).

However, with nonadditive smoothing functions, the kernel function $k(\bdx, \bdx')$ is usually a nonlinear function of multivariate $\bdx$, such as the Gaussian kernel function. %, $\exp(-\rho\|\bdx-\bdx'\|^2)$, with scale parameter $\rho$.
In a model with such a kernel function, the response can no longer be expressed in terms of additive function components and no sparse function components are available. Therefore variable selection for recovering the sparsity of $\bdx$ within the nonadditive function becomes challenging. To address this issue, some Bayesian approaches of variable selection for Gaussian process models have been developed (Linkletter et al. 2006; Savitsky et al. 2011). To the best of our knowledge, no variable selection method based on the kernel machine have been established for nonadditive smoothing function models simultaneously recovering sparsity of input variables in a nonadditive smoothing function. %(during preparing this manuscript we learned Li et al. (2011) independently proposed a similar penalized Gaussian process regression model). However, they contributed the model as the LASSO type penalty and their algorithm based on the marginal likelihood of the hyperparameters may not be efficient for high dimensional case.

Thus the goal of this paper is to study this model from the different view (Nonnegative Garrotte on Kernel) and to generalize it to include different Gaussian process kernels as a new variable selection approach on kernel machine, which is able to recover sparsity of input variables in a nonadditive smoothing function.

%Automatic relevance determination (ARD) was originally formulated in the framework of neural network in the Gaussian process context (Neal 1996, MacKay 1994), which considers having a kernel function of the form $k(\bdx,\bdx')=\exp\left\{-\sum_{j=1}^p\xi_j(x_j-x_j')^2\right\}$. ARD has been used to estimate scale parameter $\xi_j$'s on features for feature selection and classifier, such as neural network and support vector machine (SVM) (Neal, 1996; Tipping, 2001; Rasmussen and Williams, 2006). The estimation of these hyperparameters $\xi_j, j=1,...,p$ reveals the sparsity of the input variables. For those $\xi_j$'s close to zero the response becomes relatively insensitive to the corresponding input variable $x_j$. %However, ARD problem is solved by using Bayesian approach and EM algorithm which are usually expensive (Krishnapuram et al., 2004; Zou et al., 2010).
Our method is motivated by automatic relevance determination (ARD), which was originally formulated in the framework of neural networks in the context of Gaussian process (Neal, 1996; MacKay, 1994) considering kernel functions of the form $k(\bdx,\bdx')=\exp\left\{-\sum_{j=1}^p\xi_j(x_j-x_j')^2\right\}$. We model the smoothing function with a general kernel function with hyperparameters $\xi_j$'s. %'s controlling the importance of the individual predictors.
By shrinking these scale parameter $\xi_j$'s, we can select the variables. In this way, our approach can be applied to either additive or nonadditive models by choosing different $K$ structure. %our approach can be seen as a nonlinear version of a nonnegative garrote on knernel machine method (NGK).
%Since our approach can view as a nonlinear version of a nonnegative garrote on knernel machine method,
%In the rest of this paper we will refer our approach as ``nongegative garrote on kernel machine'' (NGK), which is further explained in Section 2.2.
% We further show that the variable selection problem with nonparametric models can be considered as a special case of the the problem of Learning the Kernel function via regularization (Micchelli and Pontil 2005; Lanckriet et al, 2005). From this point of view, specific kernel functions are determined by specific set of scaling parameters $\xi_j$, and by varying values of $\xi_j$'s, we optimize the objective function through learning the kernel.

For theoretical understanding of our approach, we develop the incoherence conditions.
%Ravikumar et al. (2009) studied sparsistency properties of SAMs and Bach (2008) extended the consistency conditions to MKL. So far, there are only a few theoretical studies on the consistency of linear nonnegative garrote (Yuan and Lin, 2007). Unlike LASSO where the correlation of the predictors determines the incoherence condition, the one of NGK are determined by the correlation of the vectors of the first derivatives of the smoothing function or the nonlinear components respecting to the scale parameters. This difference provides new approach to recover the sparsity of predictors when linear LASSO fails.
We show that under certain conditions sparsistency can be established. To recover sparsity of $\xi_j$'s, an efficient coordinate descent/backfitting algorithm has been developed to achieve the regularization path for $\xi_j$'s. %Our backfitting algorithm can also be applied in least squares kernel machine estimation.
%In the rest of this paper we will refer our approach as ``nongegative garrote on kernel machine'' (NGK), which is explained in Section 2.2.

The remainder of this paper is organized as follows. In Section \ref{sec2}, we first define the optimization function of our approach on the kernel model and discuss the connection of our approach with the linear nonnegative garrote model and also with the kernel machine learning problem. In Section \ref{sec3} we propose our coordinate descent updating algorithm for the solution path of the scaling parameters. In Section \ref{sec4} we discuss the necessary and sufficient conditions for consistency and sparsistency. We also show the asymptotic properties of our method with the consistent initial kernel function coefficients. In Section \ref{sec5}, we present several simulation examples. In Section \ref{sec6}, we apply our method to two real datasets: a cryptography dataset and a genetic pathway dataset. Section \ref{sec7} contains concluding remarks.

\section{Flexible Multivariate Nonparametric Modeling}\label{sec2}

\subsection{ Multivariate Nonparametric Model Using Kernel Machine}\label{sec2.1}
Consider an $n$-observation and $p$-predictor dataset $(\mby, X)$, where $X=[\mbx_1,\mbx_2, ..., \mbx_p]$ and $\mbx_j=(x_{j1}, ..., x_{jn})^T$ is an $n\times1$ vector for the $j$th predictor, $j=1,...p$.  In other words, $X=[\bdx_1,\bdx_2, ..., \bdx_n]^T$ where $\bdx^T_i$ is a $1\times p$ vector of predictors of $i$th observation, $i=1,...n$.

According to the Representer Theorem, the nonparametric multiple regression model can be expressed as (Kimeldorf and Wahba, 1971)
\be\label{e1}
\mby=\mathbf f(X)+\bel=K\bal+\bel,
\ee
where $\bel\sim N(\mathbf 0, \sigma^2I)$ and $K$ is the kernel matrix corresponding to the function space $\mathcal{H}_K$, $f\in \mathcal{H}_K$, also known as a ``Gram matrix'' of the kernel function $k(\bdx, \bdx')$. Thus the nonlinear function $f(\bdx)$ can be expressed as $\sum_{i=1}^n\alpha_ik(\bdx_i, \bdx)$, where $\bal=(\alpha_1,...,\alpha_n)^T$ is an $n\times1$ vector of the coefficients of the kernel function. Note that in model (\ref{e1}), $\mby$ is centered, i.e. $\sum y_i=0$. %, or we can assign an intercept term $\mu$ to model (\ref{e1}), which can be easily estimated by regular maximum likelihood estimation (MLE) or restricted maximum likelihood estimation (REML) method.
We also standardize $X$ such that $\sum_{l=1}^nx_{jl}=0$ and $\sum_{l=1}^{n}x_{jl}^2=1, j=1,...,p$.

To estimate $\bal$ in (\ref{e1}), least squares kernel machine estimation minimizes the least squares error with penalized norm $\|\mathbf f\|^2_{\mathcal{H}_K}=\bal^TK\bal$ induced by the kernel of the function space $\mathcal{H}_K$,
\be\label{e2}
{1\over2}\left\|\mby-K\bal\right\|^2+{1\over2}\lambda_0\bal^TK\bal,
\ee
and the solution is
\be\label{e3}
\hat{\bal}=(\lambda_0 I+K)^{-1}\mby,
\ee
where $\lambda_0>0$ is a smoothing parameter which balances the tradeoff between goodness of fit and smoothing the curve
or high dimensional surface.

\subsection{Nonnegative Garrotte on Kernel (NGK)}\label{sec2.2}
The Gram matrix can be viewed as applying a componentwise function on the similarity matrix among observations. The similarity metric between $\bdx$ and $\bdx'$ can be either the negative squared Euclidean distance, $-\|\bdx-\bdx'\|^2$, or the angle (dot product), $\bdx^T\bdx'$.   Both of the similarity metrics can be written in an additive form in terms of $p$ predictors, i.e. $-\|\bdx-\bdx'\|^2=-\{(x_1-x_1')^2+...+(x_p-x_p')^2\}$ and $\bdx^T\bdx'=x_1x_1'+...+x_px_p'$. By this additivity, the kernel matrix can be expressed as either a linear or nonlinear function of the additive form. For example, the Gram matrix by the dot product $\bdx^T\bdx'$ among observations is the linear polynomial kernel
\[\label{e4}
K(X)=\rho XX^T=\sum_{j=1}^p\rho\mbx_j\mbx_j^T=\sum_{j=1}^p\rho D^j,
\]
where $D^j=\mbx_j\mbx_j^T$, with $(k,l)$th entry $d^j_{kl}=x_{jk}x_{jl}, 1\le k, l\le n$, and $\rho$ is a scale parameter. Unlike a linear kernel, the Gaussian kernel can be expressed in a nonlinear function form because the Gram matrix with entries produced by the exponential function of $-\|\bdx-\bdx'\|^2$ is
\[\label{e5}
K(X)=\exp\left(\rho\sum_{j=1}^pD^j\right),
\]
where $D^j$ is the matrix with $(k,l)$th entry $d^j_{kl}=-(x_{jk}-x_{jl})^2$ and the $(k,l)$th entry of matrix $\sum_{j=1}^pD^j$ is $-\|\bdx_k-\bdx_l\|^2=-\sum_{j=1}^p(x_{jk}-x_{jl})^2$.

More generally, let us consider a nonnegative scale parameter $\boldsymbol \xi=(\xi_1,...,\xi_p)$ with $\xi_j$ corresponding to each predictor $\mbx_j$. Then both kernels can be expressed as
\be\label{e6}
K(\bxi,X)=g\left(\sum_{j=1}^p\xi_jD^j\right),
\ee
where $\xi_j\geq0, j=1,...,p$, and function $g(\cdot)$ is a componentwise function of matrix entries. That is, for  a linear polynomial kernel, all $\xi_j=\rho$ and $g(\cdot)$ is the identity function, and for a Gaussian kernel, all $\xi_j=\rho$ and $g(\cdot)=\exp(\cdot)$.  Note that we do not need extra constraints on the $\xi_j$'s such as $\sum\xi_j=1$. This is because for a Gaussian kernel, $\exp(\cdot)$ already places constraints on the $\xi_j$'s and, for a linear polynomial, the solution paths of the $\xi_j$'s with and without constraints differ only by a scalar factor and their sparsity properties remain the same. Thus for computational convenience we do not apply constraints on $\bxi$ for either kernels.

By introducing such nonnegative parameters to the kernel matrix, we can develop a variable selection approach for the nonparametric regression model (\ref{e1}) similar to the linear nonnegative garrotte method (Breiman, 1995). That is, we apply an extra penalty on $\bxi$ such that optimization problem (\ref{e2}) is subject to $\xi_j\geq0$ and $\sum\xi_j\leq c$ ($c$ is a positive real number), which results in the optimal problem
\be\label{e7}
{1\over2}\left\|\mby-K\boldsymbol(\bxi, X)\bal\right\|^2+{1\over2}\lambda_0\bal^TK(\bxi, X)\bal+n\lambda\sum\xi_j,
\ee
where $\lambda>0$ is a tuning parameter. We can refer to this method as ``nonnegative garrote on kernel machine''. %{\bf In kernel machine community, the relevant variable selection by the sparse hyper parameter $\bxi$ is not new, which is discussed as ARD. Most works on ARD are limited to Bayesian or EM approach. However our nonnegative garrote approach is more generalized approach so that our model can connects the ARD problem. We further provides an efficient algorithm for nonnegative garrote approach.}

\subsection{Connection with Linear Nonnegative Garotte Estimator}\label{sec2.3}
Introduced by Breiman (1995) for variable selection with the linear model $\mby=\mathbf f+\bel=X\bbe+\bel$, the linear nonnegative garotte estimator for the shrinking factor $\bxi=(\xi_1,..., \xi_p)^T$ is the solution that minimizes
\be\label{e8}
{1\over2}\left\|\mby-\sum_{j=1}^p\xi_j\mbx_j\tilde\beta^{OLS}_j\right\|^2+n\lambda\sum\xi_j,\;\hbox{subject to } \xi_j\ge0, \forall j,
\ee
where $\tilde\beta^{OLS}_j$ is the initial estimate of $\beta_j$ using ordinary least squares.  For an orthonormal design $X^TX=I$ (i.e. $\mbx_i^T\mbx_j=\delta_{ij}$ such that $\delta_{ij}=0\,\forall i\ne j$ and $\delta_{ij}=1\,\forall i=j$), the nonnegative garrote solution for $\beta_j$ is
\[\label{e9}
\hat\beta_j=\hat\xi_j\tilde\beta_j^{OLS}\hbox{ and } \hat\xi_j=\left(1-{n\lambda\over \left(\tilde\beta_j^{OLS}\right)^2}\right)_+,\;j=1,...p,
\]
where subscript ``$+$'' indicates the positive part of the expression. We can see that (\ref{e8}) is a special case of (\ref{e7}) with a linear polynomial kernel. To see this, consider (\ref{e7}) without penalty on $\bal$. Thus $\lambda_0=0$ and kernel matrix $K=\sum\xi_j\mbx_j\mbx_j^T$. Then the least squares kernel machine solution is related to the OLS solution by choosing the initial $\tilde{\bal}=\mby$ and we obtain the initial estimate for the response
\[\label{e10}
\tilde{\mathbf f}=K\tilde{\bal}=\sum\xi_j\mbx_j\mbx_j^T\mby=\sum\xi_j\mbx_j\tilde{\beta}_j^{OLS}=\sum\xi_j\tilde f_j,
\]
where $\tilde f_j$ represents the initial marginal response and the OLS estimate of the linear model is $\tilde{\beta}_j^{OLS}=\mathbf e_j^T(X^TX)^{-1}X^T\mby=\mbx_j^T\mby$ since $X^TX=I$. $\mathbf e_j^T$ here is the selection vector with 1 in the $j$th position.

%Yuan and Lin (2007) suggested that the linear nonnegative garrote can be used in combination with the initial estimator of $\bbe$ other than least squares estimate.
Yuan and Lin (2007) proposed a general linear nonnegative garrote approach
\be\label{e11}
{1\over2}\left\|\mby-Z\bxi\right\|^2+n\lambda\sum\xi_j,\;\hbox{subject to } \xi_j\ge0, \forall j,
\ee
where $Z$ is the matrix of columns of initial marginal response estimates. They proved that if the initial estimate is consistent, then the nonnegative garrote estimate is also consistent. Based on this idea, Yuan (2007) applied the nonnegative garrote component selection method to functional ANOVA models using the initial estimates of the function components $\tilde f_j$ as columns of $Z$ ($\tilde f_j$ can be an interaction component). %So the linear nonnegative garrote method can be applied to additive functional ANOVA models with interaction components.

In Section \ref{sec4}, we will further derive similar approximated linear form of objective function (\ref{e7}) and show that $Z$ is a matrix with columns $\left({{\partial K}\over{\partial \xi_j}}\right)\tilde\bal=D^j\tilde\bal, j=1,...,p$. In this case, it requires a local linear approximation of the kernel, say $K(\bxi)\approx K(\bxi^*)+\sum_{j=1}^p(\xi_j-\xi_j^*)K'_j(\bxi^*)$. For a general kernel function, $\left({{\partial K}\over{\partial \xi_j}}\right)\tilde\bal$ can be understood as the slope of the change of initial $\tilde f$ along $\xi_j$ direction given initial $\tilde\bal$.  However, for nonlinear kernel functions such as the Gaussian kernel, we can not derive the algorithm with the linear form of (\ref{e11}).

%, which is usually a too strong approximation to derive an algorithm that converges. On the other hand, (\ref{e11}) can be derived without such approximation for linear polynomial kernel.

\subsection{Connection with Kernel Machine Learning}\label{sec2.4}
First define function $Q_{0}(\cdot)$ as
\be\label{e12}
Q_0\left(K(\bxi, X), \bal\right)={1\over{2}}\left\|\mby-K\boldsymbol(\bxi, X)\bal\right\|^2+{\lambda_0\over{2}}\bal^TK(\bxi, X)\bal.
\ee
The following lemma shows that $Q_0$ can be expressed in a form of function of $K$.

\noindent{\bf Lemma 1:}\label{t1} {\it If $K$ is in the set of all kernels on input set $\mathbb{X}$, and if for a set of distinct points $X\in\mathbb{X}$ $K(X)$ is positive definite, then
\be\label{e13}
Q_0(K)={\lambda_0\over{2}}\mby^T\left(\lambda_0I+K\right)^{-1}\mby,
\ee
and $Q_0(K)$ is a non-increasing convex function of $K$.}

The form of (\ref{e13}) can be easily derived since the solution for the least squares kernel machine problem is $\hat{\bal}=(\lambda_0I+K)^{-1}\mby$. Plug this solution back into (\ref{e12}) and by simple algebra we obtain (\ref{e13}). A formal proof of Lemma 1 and convexity of $Q_0(K)$ can be found in Micchelli and Pontil (2005) and Lanckriet et al. (2004). Then the solution $\hat\bxi$ of the optimization problem (\ref{e7}) is considered as
\be\label{e14}
\begin{split}
\min_{K}\; Q_0(K)&={\lambda_0\over{2}}\mby^T\left(\lambda_0I+K\right)^{-1}\mby,\\
\textrm{subject to}\; K\in\mathbb{K}^*&=\left\{K(\bxi, X): \bxi\in \mathbb{R}^p_+\hbox{ and }\sum_{j=1}^p\xi_j\le c, j=1,...,p\right\},
\end{split}
\ee
where $\mathbb{R}^p_+$ is set of $p$ dimensional nonnegative real numbers.

The objective function (\ref{e14}) implies that we have a kernel based learning problem on $\mathbb{K}^*$, a subset of all kernels on input set $\mathbb{X}$. More generally, the problem associated with the function $Q_0(K)$ and the kernel $K$ is the variation problem (Micchelli and Pontil, 2005)
\be\label{e15}
Q_0(\mathbb{K})=\inf\{Q_0(K): K\in\mathbb{K}\},
\ee
where $\mathbb{K}$ is a convex set of all positive semidefinite kernel functions. Thus our problem can be viewed as a special case of (\ref{e15}), as learning the kernel function via regularization ${\lambda_0\over{2}}\|\mathbf f\|_{\mathcal{H}_{K}}^2={\lambda_0\over{2}}\bal^TK\bal$, subject to $K\in\mathbb{K}^*$, where $\mathbb{K}^*\subset\mathbb{K}$. If optimizing $Q_0$ on $\mathbb{K}$, this is the problem of learning the kernel discussed by Micchelli and Pontil (2005).

Although our problem is learning the kernel $K\in\mathbb{K}^*$, it is different from the problem of learning the kernel function discussed by Micchelli and Pontil (2005), Lanckriet et al. (2004), and Rakotomamonjy et al. (2008). This is because the set $\mathbb{K}^*\subset\mathbb{K}$ is usually not convex and the optimization problem is learning the kernel through a nonlinear function $g(\cdot)$ on $\bxi$.

We can express (\ref{e7}) as a function of $\bxi$,
\be\label{e16}
Q(\bxi)=Q_0(\bxi)+\lambda\sum_{j=1}^p\xi_j={\lambda_0\over{2}}\mby^T\left(\lambda_0I+K(\bxi)\right)^{-1}\mby+\lambda\sum_{j=1}^p\xi_j.
\ee
The convexity of $Q_0(\bxi)$ is interesting since it determines the convexity of $Q(\bxi)$ and convex objective functions have many convenient properties in the optimization problem. Unfortunately, however, it is not straightforward to determine the convexity of $Q_0(\bxi)$ as its convexity completely depends on the kernel function $K(\bxi)$ and $X$.
The following Lemma shows a sufficient condition for $Q (\bxi)$ to be a convex function of $\bxi$.

\noindent{\bf Lemma 2:} {\it If matrix set $K(\bxi)=g(\sum_{j=1}^p\xi_jD^j)$ is concave on $\bxi$, i.e. $K(\theta\bxi+(1-\theta)\bxi')\succeq\theta K(\bxi)+(1-\theta)K(\bxi')$ where $0\leq\theta\leq1$, then the regularization problem (\ref{e16}) is convex on $\bxi\in\mathbb{R}^p_+$.}

This can be easily shown by the composition theorem (Boyd and Vandenberghe, 2004). That is, for a function $f(x)=h\{g(x)\}$, $f$ is convex if $h$ is convex and non-increasing and $g$ is concave (see Appendix \ref{apen1}). An obvious example of a concave kernel $K$ is the linear polynomial kernel (which is convex too). So we conclude that the objective function $Q_0(\bxi)$ is a convex function of $\bxi$ for a linear polynomial kernel and $Q(\bxi)=Q_0(\bxi)+\lambda\sum\xi_j$ is a strictly convex function of $\bxi$. Thus the regularization problem (\ref{e16}) has many of the nice properties of convex optimization. In particular $Q(\bxi)$ is strictly convex, the solution $\hat\bxi$ is unique.

However, in many cases, it is not straightforward to derive the concavity or convexity of $K(\bxi)$. For instance, with the Gaussian kernel, since $K(\bxi)$ is neither concave nor convex on $\bxi$, it is difficult to determine the convexity of $Q(\bxi)$. The most ideal scenario is, $Q(\bxi)=Q_0(\bxi)+\lambda\sum\xi_j$ is quasicovex (unimodal) so that $Q(\bxi)$ has a unique minimum. In this case, the Hessian matrix of $Q(\bxi)$ may not be positive (semi-)definite everywhere. However, one can expect a positive (semi-)definite Hessian matrix in the neighborhood of a minimum, that is, $H=\left\{{\partial^2Q(\bxi)}\over{\partial\bxi^T\partial\bxi}\right\}\succeq0$.

In practice, we start with an initial $\hat\bxi$ close to zero when solving the optimization problem. Hence we can assume the initial values and the solution path are always in the neighborhood of a minimum where $H\succeq0$. This is a reasonable assumption with Gaussian kernels since in the sparsity problem most $\xi_j$'s are zero and, in general, non-zero $\xi_j$'s are all small positive numbers. When the $\xi_j$'s are small numbers or zeros,  Gaussian kernel can be well approximated to the linear order of the Taylor expansion as a concave matrix of $\bxi$. In the following section we provide the regularity conditions which are usually satisfied in least squares error optimization.

\subsection{Some Notation and Regularity Conditions}\label{sec2.5}
We first define some notation. Let $\bxi^*$ and $\hat{\bxi}$ represnt the true $\bxi$ and minimum solution of (\ref{e7}), respectively. Suppose vector $\bxi^*$ is sparse, i.e. some $\xi_j^*=0$. Without loss of generality, denote $\bxi^*=(\xi^*_1,...,\xi^*_p)^T=\left({\bxi^*_1}^T,{\bxi^*_0}^T\right)^T$, where $\bxi^*_1$ is the vector of the first $a$ nonzero $\xi^*_i$'s and $\bxi^*_0$ is the zero vector. Define the nonzero index set of $\bxi^*$ as $\ma:=\{j\in\{1,...,p\}|\xi^*_j>0\}$, and denote $\hat{\ma}:=\{j\in\{1,...,p\}|\hat\xi_{j}>0\}$ as the nonzero index set of $\hat\bxi$. Note that $\mathcal{A}$ has relatively small cardinality $a=|\ma|$, the number of true nonzero $\xi_j$'s.

Let the least squares error estimate of $\bal$ be $\hat\bal=\Delta^{-1}(\hat\bxi)\mby$, where $\Delta(\bxi)=\lambda_0I+K(\bxi)$, and denote the true $\bal$ vector as $\bal^*$. For a given $\bal^*$ and estimate $\tilde\bal$ we define the following matrices and their respective partitions:
\be\label{e18}
Z=[\mathbf z_1,...,\mathbf z_p]=[Z_1, Z_0]=\left[\left\{K_j'(\bxi^*)\bal^*\right\}_{1\le j\le a},\left\{K_j'(\bxi^*)\bal^*\right\}_{a+1\le j\le p}\right]
\ee
\be\label{e19}
\tilde Z=[\tilde{\mathbf z}_1,...,\tilde{\mathbf z}_p]=[\tilde Z_1,\tilde Z_0]=\left[\left\{K_j'(\bxi^*)\tilde{\bal}\right\}_{1\le j\le a},\left\{K_j'(\bxi^*)\tilde{\bal}\right\}_{a+1\le j\le p}\right],
\ee
where $K_j'(\bxi^*)=\left.{{\partial K}\over{\partial\xi_j}}\right|_{\bxi^*}$, obtained by taking the partial derivative of the componentwise entries of $K$.
Note that $K_j'\bal$ is an $n\times 1$ vector and the $Z$'s and $\tilde{Z}$'s are $n\times p$ matrices. Some covariance matrices are also defined as
\be\label{e20}
\begin{array}{lcclcc}
\Sigma_{11}&=&\left(n^{-1}Z_1^TZ_1\right),& \Sigma_{01}&=&\left(n^{-1}Z_0^TZ_1\right),\\
\tilde\Sigma_{11}&=&\left(n^{-1}\tilde Z_1^T\tilde Z_1\right),& \tilde\Sigma_{01}&=&\left(n^{-1}\tilde Z_0^T\tilde Z_1\right),\\
\end{array}
\ee
where $\Sigma_{11}$ and $\tilde\Sigma_{11}$ are assumed to be invertible.

We then further define the $p\times1$ vector $\mathbf v_n(\bxi)$ and $p\times p$ matrix $M_n(\bxi)$ as follows:
\[\label{e21}
\mathbf v_n(\bxi)=\lambda_0^{-1}n^{-1/2}{{\partial Q_0(\bxi)}\over{\partial\bxi}}=-\left\{{1\over{2\sqrt{n}}}\mby^T\Delta^{-1}\left({{\partial K}\over{\partial\xi_j}}\right)\Delta^{-1}\mby\right\}_{1\le j\le p}^T,
\]
\[\label{e22}
M_n(\bxi)=(\lambda_0n)^{-1}{{\partial^2 Q_0(\bxi)}\over{\partial\bxi^T\partial\bxi}}=\left\{{1\over{2n}}\mby^T\Delta^{-1}\left[{{\partial K}\over{\partial\xi_i}}\Delta^{-1}{{\partial K}\over{\partial\xi_j}}+{{\partial K}\over{\partial\xi_j}}\Delta^{-1}{{\partial K}\over{\partial\xi_i}}-{{\partial^2K}\over{\partial\xi_i\partial\xi_j}}\right]\Delta^{-1}\mby\right\}_{1\le i,j\le p}.
\]
These are analogous to the negative score and Hessian matrix of a log likelihood function $-Q_0(\bxi)$, respectively. To see this and the regularity conditions,  we first consider the log likelihood function of our model. %From a Bayesian stand point, assume
%\bse\label{e23}
%\mby|\bal,\bxi&\sim& N\left(K(\bxi,X)\bal,\sigma^2I\right),\\
%\bal|\bxi&\sim& N\left(\mathbf 0,\sigma^2_\alpha K(\bxi,X)^{-1}\right),\\
%\xi_j&\sim& \lambda e^{-\lambda\xi_j}, j=1,...,p.
%\ese
Then, up to an additive constant, the negative log likelihood function of $\bxi$ is
\[\label{e24}
{1\over{2\sigma^2}}\left\|\mby-K\boldsymbol(\bxi, X)\bal\right\|^2+{1\over{2\sigma^2_\alpha}}\bal^TK(\bxi, X)\bal-{1\over2}\log|K(\bxi,X)|+n\tilde\lambda\sum\xi_j.
\]
Letting $\lambda_0={\sigma^2\over\sigma^2_\alpha}$ and $\sigma^2\tilde\lambda=\lambda$, the above expression is equivalent to
\be\label{e25}
{1\over{2}}\left\|\mby-K\boldsymbol(\bxi, X)\bal\right\|^2+{\lambda_0\over{2}}\bal^TK(\bxi, X)\bal-{\sigma^2\over2}\log|K(\bxi,X)|+n\lambda\sum\xi_j.
\ee
Expression (\ref{e25}) only differs from (\ref{e7}) with respect to a $\log|K|$ term. Thus, strictly speaking, estimating $\hat\bxi$ by (\ref{e7}) no longer provides the MLE estimate. We omit the $\log|K|$ term since in the NGK model, values of the $\xi_j$'s are usually sparse and small. Hence, in the region where we estimate $\hat\bxi$, and the determinant of $K$ is almost constant of $\bxi$ for both the Gaussian kernel and linear polynomial kernel. In this sense, (\ref{e7}) and (\ref{e25}) are equivalent. However, our algorithm benefits greatly from omitting $\log|K|$ since derivatives of this term result in complicated expressions. Therefore we assume the minimum of (\ref{e25}) is not greatly affected by $\log|K|$ and we can still consider our objective functions $Q(\bxi)$ and $Q_0(\bxi)$ as the log likelihood function of $(\bal,\bxi)$ with and without a prior for $\bxi$, respectively.

Based on these arguments, considering the convexity and differentiability of $Q_0(\bxi)$, we can assume that the regularity conditions of the log likelihood also apply to $Q_0(\bxi)$ such that
\be\label{e26}
\begin{split}
\lim_{n\rightarrow\infty}&\mathbf v_n(\bxi^*)\rightarrow\mathbf v^*,\;\hbox{and }\|\mathbf v^*\|_\infty<\infty,\\
\lim_{n\rightarrow\infty}&M_n(\bxi^*)\rightarrow M^*,\;\hbox{and } \|M^*\|_\infty<\infty.
\end{split}
\ee
Particularly
\be\label{e27}
\lim_{n\rightarrow\infty}\mathbf v_n(\bxi^*)\rightarrow-\lim_{n\rightarrow\infty}\left\{{1\over{2\sqrt{n}}}{\bal^*}^TK'_j(\bxi^*)\bal^*\right\}_{1\le j\le p}^T=\mathbf v^*.
\ee
These regularity conditions indicate that $\mathbf v_n(\bxi)=O_p(1)$ at $\bxi^*$ and $M_n(\bxi)$ is finite and positive (semi-)definite at $\bxi^*$. These conditions are consistent with the convexity assumption of $Q_0(\bxi)$ discussed in the previous section.

\section{Methodology}\label{sec3}
In this section, we provide an efficient algorithm to solve the objective function (\ref{e7}) with a given initial $\hat\bal$.
\subsection{Backfitting Algorithm to Update $\bxi$}\label{sec3.1}
An efficient algorithm to achieve the regularization path of (\ref{e7}) for $(\hat{\bal}, \hat{\bxi})$ is still an open problem.  One possible approach is to iteratively update between $\hat{\bal}$ and $\hat{\bxi}$ until convergence from an initial $(\hat{\bal}^{(0)}, \hat{\bxi}^{(0)})$. This updating approach, however, could be very expensive and may not be able to converge. Another possible approach is the one-step update algorithm proposed by Lin and Zhang (2006) for COSSO. That is, at each fixed $\lambda$, solve $\hat\bxi$ with given $\hat\bal$, then update $\hat\bal=\Delta^{-1}\mby$ with the new $\hat\bxi$ and continue to the next step. However, the one-step update algorithm may not be necessary to solve the solution path of $\hat\bxi$ as long as we have some consistent initial estimate of $\tilde\bal$ and keep it fixed through the entire solution path. In Section \ref{sec4} we will show theoretically that, as long as $\tilde\bal$ is consistent, the sparsity of $\bxi$ can be recovered as $n$ increases. Although the initial $\bal$-fixed algorithm may not results in a consistent estimation of $\bxi$, we will also show under certain conditions that the estimation consistency of $\bxi$ can be achieved.

With fixed initial consistent $\tilde\bal$, the algorithm to update $\bxi$ becomes efficient. The  algorithm we propose in the following can be viewed as the non-linear version of the coordinate descent algorithm for nonnegative garrotes. In special cases where linear polynomial kernels or other additive multiple kernels are considered, our algorithm is equivalent to the least angle regression selection (LARS) algorithm.

The steps of our algorithm are summarized as follows:
\begin{enumerate}[$\bullet\,${\it Step} 1]
\item Initialize $\tilde\bal=(\lambda_0I+K(\rho))^{-1}\mby$ and $\lambda_0=\sigma^2/\sigma^2_\alpha$ by setting all $\xi_j=\rho$ and fitting the least squares kernel machine by MLE/REML methods.
\item Determine the initial $\lambda$ for which all $\hat\xi^{(0)}_j=0$
\[\label{e28}
\lambda^{(0)}=\max_{j}\left\{n^{-1}\left(\tilde{\mby}-K(0)\tilde\bal\right)^T\left(K'_j(0)\tilde\bal\right)\right\},
\]
where $\tilde{\mby}=\mby-{\lambda_0\over2}\tilde\bal$.
\item\label{s1} Update $\hat\bxi$ coordinate wise at $\lambda^{(k+1)}$ with given $\tilde\bal$ by the following equation until converge:
\be\label{e29}
\hat\xi_j=\left[\tilde\xi_j+{{(\tilde{\mby}-K\tilde\bal)^TK'_j\tilde\bal-n\lambda^{(k+1)}}\over{\left(K'_j\tilde\bal\right)^T\left(K'_j\tilde\bal\right)}}\right]_+,
\ee
where $\tilde\xi_j$ denotes the previously updated $\hat\xi_j$, and $K$ and $K'_j$ are calculated from previously updated $\tilde\xi_j$'s.

\item Decrease $\lambda$ and repeat step \ref{s1}.
\item Stop when model selection criterion reaches minimum or last $\lambda=0$.
\end{enumerate}
%In Section \ref{sec4}, we will consider the linear approximation assumption of kernel to derive consistency conditions for $\hat\bxi$. However, in practice this assumption is too strong to develop efficient algorithm to update $\hat\bxi$.

In Step 3, we derive the updating equation (\ref{e29}) for $\hat\bxi$ via an approximation of $K(\bxi)$. At given $\lambda$, assuming the current iteration $\tilde\bxi$ is close to the minimum solution $\hat\bxi$, the kernel matrix can be extended in one coordinate direction around $\tilde\xi_j$:
\[\label{e30}
K(\tilde\bxi_{-j},\hat\xi_j)=K(\tilde\bxi)+(\hat\xi_j-\tilde\xi_j)\left({{\partial K}\over{\partial\xi_j}}\right)_{\tilde\bxi}+O(\|\hat\xi_j-\tilde\xi_j\|^2),
\]
where $(-j)$ denotes exclusion of $\xi_j$. In simple notation
\be\label{e31}
K(\tilde\bxi_{-j},\hat\xi_j)\approx K+(\hat\xi_j-\tilde\xi_j)K'_j,
\ee
where $K=K(\tilde\bxi)$ and $K'_j=\left({{\partial K}\over{\partial\xi_j}}\right)_{\tilde\bxi}$. As an example: for a Gaussian kernel $K'_j=K\circ D^j$ and  for a linear polynomial kernel $K'_j=D^j$, where ``$\circ$'' denotes the Schur product or entrywise product of two matrices.

The updating solution of $\hat\xi_j$ given $\tilde\bxi$ is achieved by plugging (\ref{e31}) into (\ref{e7}) and solving $\hat{\xi}_j=\arg\min{(\ref{e7})}$ given $\tilde\bal$ and $\lambda_0$. We notice that expression (\ref{e29}) is similar to the backfitting algorithm (Ravikumar et al., 2009; Hastie and Tibshirani, 1990) in nonparametric additive models, except our algorithm is a version of backfitting on nonadditive models by considering the $\xi_j$ updating step as

\begin{enumerate}[$\bullet\,${\it Step} a]
   \item Initialize $\hat\xi_j=\hat\xi_j^{(k)}, j=1,...,p$ with $\tilde\bal$ given.
   \item \label{s2}
\begin{enumerate}[(1)]
  \item Compute the residual, $\mathbf r_j=\tilde{\mby}-K\tilde\bal+\tilde\xi_jK'_j\tilde\bal$.
  \item Project the residual onto $\mathbf z_j=K'_j\tilde\bal$ and $P_j=\mathbf z_j^T\mathbf r_j$.
  \item Update $\hat\xi_j=\left({{P_j-n\lambda}\over{\|\mathbf z_j\|}^2}\right)_+$.
\end{enumerate}
\item Repeat \ref{s2} until the individual $\hat\xi_j$'s do not change.
\end{enumerate}

\subsection{Advantages of our  Algorithm}\label{sec3.2a}
When we use a linear polynomial kernel, $K=\sum\xi_j\mbx_j\mbx_j^T$, we revert back to the additive kernel case, $K=\sum\xi_jK_j$, which has been thoroughly discussed by Bach (2008) and Rakotomamonjy et al. (2008) for the multiple kernel learning (MKL) with LASSO. For the additive kernel case, this is closely related to functional ANOVA models. Yuan and Lin (2007) proposed nonnegative garrote component selection in these models, and introduced the LARS algorithm for solving the linear nonnegative garrote problems when $p<n$. When $p>n$, the LARS algorithm may not work well due to singularity of the active variable correlation matrix and reliance on generalized inverse matrices. On the other hand, our algorithm has two main advantages: first it works for $p>n$, second it works with nonadditive
kernels. In addition, our algorithm is related to ARD, which is typically discussed in a Bayesian context (Neal, 1996; Krishnapuram et al., 2004; Zou et al., 2010). To the best of our knowledge, our algorithm is the only non-Bayesian approach for determining the hyper parameters in ARD with penalty on $\bxi$, and has the added bonus of being more efficient.

\subsection{Model Selection}\label{sec3.2}
Variable selection depends on how to select the penalty parameter. After determining the penalty parameter by some criteria, one can decide the final model which contains the most relevant variables to the response. However, as we discussed before, NGK variable selection is rather new topic within the kernel machine framework. There is no similar work available to provide a perfect criterion for selecting penalty parameter $\lambda$. One can choose a optimal $\lambda$ at the minimum of the criterion and then obtain variables in a model at this optimal $\lambda$. However in practice most of the popular criteria such as BIC, Cp and GCV may become flat. It becomes hard to determine an appropriate minimum. Another issue is that there is no perfect criterion in existence.  The performance of any criterion not only depends on the model, but also depends on the data structure. Hence in our study, we propose to select variables according to the selection probability or frequency of individual variables. This probability is achieved by some resampling procedures with variable selected by least squares kernel machine BIC for each single resampling. We propose two resampling procedures: one is based on bootstrapping for large sample size and the other is based on permutation for small sample size. Our resampling procedures are further described in Section \ref{sec6.1} and \ref{sec6.2}, respectively.

The least squares kernel machine BIC is defined as
 \[\label{e32}
BIC=\log(RSS)+\frac{df\,\log(n)}{n},
\]
where $RSS=(\mby-\hat{\mathbf f})^T(\mby-\hat{\mathbf f})$. For given minimum solution $\hat\bxi$, the estimated function $\mathbf f$ can be expressed as $\hat{\mathbf f}=S\mby$, where $S$ is the smoothing matrix. For the least squares error kernel machine, $S=K(\hat\bxi)\left(\lambda_0I+K(\hat\bxi)\right)^{-1}$, the degrees of freedom of the kernel machine smoother $S$ is defined as $df=\hbox{Trace}(S)$. BIC was used by Liu et al. (2007) in the semiparametric mixed model with the least squares kernel machine.

\section{Some Theoretical Properties}\label{sec4}
%In recent years, tremendous progress has been made to understand the mechanism of variable selection in linear LASSO. Those theoretical works focus on consistency analysis of LASSO estimation.
Consistency in variable selection problem includes two aspects: estimation consistency and model selection consistency.  Between the two, one does not necessarily imply the another. The former requires $\hat\bxi-\bxi^*\rightarrow \mathbf 0$ as $n\rightarrow \infty$, and the later requires $\lim_nP(\hat{\ma}=\ma)\rightarrow1$. The model consistency is also called sparsistency, shorthand for ``sparsity pattern consistency'' (Ravikumar et al., 2009).

In this section, similar to consistency of LASSO, we first show that under certain conditions the NGK estimator is $\sqrt{n}$ consistent. Then we will further discuss conditions for which the NGK estimators are sparsistent for initial $\tilde\bal$. This is important because, in our NGK algorithm, we assume the initial $\tilde\bal$ is fixed.

\subsection{Necessary and Sufficient Conditions for Consistency of $\hat{\bxi}$}\label{sec4.1}
We first establish the $\sqrt{n}$ consistency of $\hat\bxi$ estimation in Theorem 1 and then provide sufficient and necessary conditions in Lemma 3.

\noindent{\bf Theorem 1:} {\it Under the regularity conditions (\ref{e26}), if $\sqrt n\lambda\rightarrow0$, then there exists a local minimum $\hat{\bxi}$ of $Q(\bxi)$ such that $\|\hat{\bxi}-\bxi^*\|=O_p(n^{-1/2})$}.

The proof of Theorem 1 is similar to Fan and Li (2001) and Wang and Leng (2007), where both used the regularity conditions of the log likelihood function. We show the proof for Theorem 1 in Appendix \ref{apen2}. The $\sqrt{n}$ estimation consistency of $\hat\bxi$ guarantees that, when $n$ is large enough, the minimum solution of (\ref{e16}) is consistent with $\bxi^*$. This Theorem means when $n$ is sufficiently large, the kernel matrix $K(\hat\bxi)$ is close to $K(\bxi^*)$.

Defining the sign function of $\bxi$,
\[\label{e36}
\sgn(\xi_j):=\left\{\begin{array}{l l}
  +1,  & \hbox{ if }\; \xi_j>0     \\
   0,  & \hbox{ if }\; \xi_j=0,  \end{array} \right.
\]
%Then similar argument has been shown for additive nonlinear regression and MKL (Ravikumar et al., 2009; Bach, 2008). Wainwright (2009) further explored those conditions for LASSO and provided sharper threshold for the convergence rate. In our model, we can show some similar results.
the following lemma states the necessary and sufficient conditions for $\hat{\bxi}$ to be consistent.

\noindent{\bf Lemma 3:} {\it Given initial $\tilde\bal=\bal^*$, the necessary and sufficient conditions for $\hat{\bxi}$ to be consistent, i.e. $\lim_nP(\hat{\ma}=\ma)=1$ or $\lim_nP\{\sgn(\hat\bxi)=\sgn(\bxi^*)\}=1$, are
\begin{subequations}
\be\label{e37a}
{1\over n}Z^T_0\left(\bel-{\lambda_0\over2}\bal^*\right)-Z^T_0Z_1(Z^T_1Z_1)^{-1}\left[{1\over n}Z^T_1\left(\bel-{\lambda_0\over2}\bal^*\right)-\lambda\mathbf 1\right]\preceq\lambda\mathbf1,
\ee
\be\label{e37b}
\bxi^*_1+\left({1\over n}Z^T_1Z_1\right)^{-1}\left[{1\over n}Z^T_1\left(\bel-{\lambda_0\over2}\bal^*\right)-\lambda\mathbf 1\right]\succ\mathbf0.
\ee
\end{subequations}}
Note that in the above expressions, $\mathbf 1$ is a vector of 1's (size different for (\ref{e37a}) and (\ref{e37b})).

To prove Lemma 3, we need some approximation form of (\ref{e7}). According to Theorem 1, $\hat\bxi$ is $\sqrt{n}$-consistent, i.e. $\hat\bxi\rightarrow\bxi^*$ as $n\rightarrow\infty$. The linear approximation of the kernel function holds: $K(\hat\bxi)= K(\bxi^*)+\sum_{j=1}^p(\hat\xi_j-\xi^*_j)K_j'(\bxi^*)+O_p(\|\hat\bxi-\bxi^*\|^2)$. Given $\tilde\bal=\bal^*$,  $\mby-\mathbf f=\mby-K(\bxi^*)\bal^*=\bel$. Plugging $\tilde\bal=\bal^*$ and approximated $K(\hat\bxi)$ into the expression of $Q(\bxi)$ in (\ref{e7}), the regularization problem is approximated as
\be\label{e38}
{1\over2}\left\|\mby-K(\bxi^*)\bal^*-\sum_{j=1}^p(\hat{\xi}_j-\xi^*_{j})K_j'(\bxi^*)\bal^*\right\|^2+{\lambda_0\over2}\sum_{j=1}^p(\hat\xi_j-\xi^*_{j}){\bal^*}^TK_j'(\bxi^*)\bal^*+n\lambda\sum_{j=1}^p\hat\xi_j.
\ee
By using notation $\tilde{\mby}=\mby-K(\bxi^*)\bal^*-{\lambda_0\over2}\bal^*+\sum_{j=1}^p\xi^*_jK_j'(\bxi^*)\bal^*=\bel-{\lambda_0\over2}\bal^*+Z\bxi^*$ and rearranging the above expression, we have the following equivalent expression
\be\label{e39}
{1\over2}\left\|\tilde{\mby}-Z\hat\bxi\right\|^2+n\lambda\sum_{j=1}^p\hat\xi_j.
\ee
Expression (\ref{e39}) is similar in form to the linear nonnegative  garrotte objective function (\ref{e11}) proposed by Yuan and Lin (2007) except for the modified response $\tilde{\mby}$ has a non-linear term ${\lambda_0\over2}{\bal^*}^T$.

Note that we use the above approximation of the kernel only for theoretical analysis purpose. For algorithm derivation, we only approximate the kernel in one $\xi_j$ direction. Thus we can not derive a form similar to (\ref{e39}) for a Gaussian kernel because $\tilde{\mby}$ and  the $Z$ matrix are no longer fixed and updated by $\hat\bxi$ each iteration (See Section \ref{sec3}). For a linear polynomial kernel, since the kernel is a linear combination of multiple kernels, no approximation is needed and we can derive the exact linear negative garrote form as above.

When the minimum solution of $\hat\bxi$ is close to $\bxi^*$ as Theorem 1 states, the solution of (\ref{e39}) is consistent with the solution of (\ref{e7}). Thus we can start from (\ref{e39}) to derive the incoherence conditions as (\ref{e37a})-(\ref{e37b}) (see Appendix \ref{apen3}).

\subsection{Recovery of Sparsity}
Note in Lemma 3, conditions (\ref{e37a})-(\ref{e37b}) are derived with the initial $\tilde\bal=\bal^*$. However, in practice, we consider $\delta$-consistent $\tilde\bal$ and $\tilde Z$ matrix. A question arises about whether or not we can similarly solve $\hat\bka_0$  and $\hat\bxi_1$ based on $\tilde\bal$,
\begin{subequations}
\be\label{e40a}
\lambda\hat\bka_0={1\over n}\tilde Z^T_0\left(\bel-{\lambda_0\over2}\tilde\bal\right)-\tilde Z^T_0\tilde Z_1(\tilde Z^T_1\tilde Z_1)^{-1}\left[{1\over n}\tilde Z^T_1\left(\bel-{\lambda_0\over2}\tilde\bal\right)-\lambda\mathbf 1\right],
\ee
and
\be\label{e40b}
\hat\bxi_1=\bxi^*_1+\left({1\over n}\tilde Z^T_1\tilde Z_1\right)^{-1}\left[{1\over n}\tilde Z^T_1\left(\bel-{\lambda_0\over2}\tilde\bal\right)-\lambda\mathbf 1\right],
\ee
\end{subequations}
such that we can use them to recover the sparsity of $\bxi^*$, where $\hat\bka_0$ is the subgradient of $\|\hat\bxi\|_1$ corresponding to those $\hat\xi_j=0$ (see the Appendix \ref{apen3}). To show these equations (\ref{e40a})-(\ref{e40b}) are satisfied, we consider additional conditions required on $\tilde\bal$ for recovering sparsity for how fast it converges to $\bal^*$.

The above argument shows that, if we have a consistent estimate of $\bal^*$, we can recover sparsity of $\bxi$ using (\ref{e40a})-(\ref{e40b}) so that we do not need to estimate $\hat\bal$ and $\hat\bxi$ at the same time. Based on this idea our algorithm is developed. In our algorithm we use some $\tilde\bal$ as the initial value and keep it fixed for the entire solution path of $\bxi$. %Similar situation has been discussed in linear nonnegative garrote by Yuan and Lin (2007).

Thus, motivated by consistency conditions (\ref{e37a})-(\ref{e37b}), we consider the following zero noise incoherence conditions on the $Z$ matrix:
\begin{subequations}
\be\label{e41a}
\Sigma_{01}\Sigma_{11}^{-1}\mathbf 1-{\lambda_0\over{2n\lambda}}Z_0^TP\bal^*\preceq(1-\gamma)\mathbf 1
\ee
where $\gamma\in(0,1]$, and $P=[I-Z_1(Z_1^TZ_1)^{-1}Z_1^T]$ is a projection matrix. Expressions (\ref{e37a})-(\ref{e37b}) and (\ref{e41a}) are calculated based on the true $\bal^*$. We can show that as long as $\tilde\bal$ is $\delta$-consistent with $\delta\rightarrow0$, the similar condition, $\tilde\Sigma_{01}\tilde\Sigma_{11}^{-1}\mathbf 1-{\lambda_0\over{2n\lambda}}\tilde Z_0^T\tilde P\tilde\bal\preceq(1-\tilde\gamma)\mathbf 1$ is satisfied (see Appendix \ref{apen4}), where $\tilde\gamma\in(0,1]$. Furthermore, we need the following assumptions for $\tilde\bal$ based calculations.
\be\label{e41b}
\Lambda_{min}(\tilde\Sigma_{11})\ge \tilde C_{min}>0
\ee
\be\label{e41c}
\tilde\Sigma_{01}\tilde\Sigma_{11}^{-1}\rightarrow\Sigma_{01}\Sigma_{11}^{-1}\hbox{ with rate no slower than }\delta,
\ee
\end{subequations}
where $\Lambda_{min}(\cdot)$ denotes the minimum nonnegative eigenvalue.\\
There are two interesting features of (\ref{e41a}). First, unlike the incoherence conditions of linear LASSO where $\Sigma_{11}$ and $\Sigma_{01}$ are the correlation matrices of predictors,  $\Sigma_{11}$ and $\Sigma_{01}$ in (\ref{e41a}) are the correlation matrices of the ${\mathbf z}_j$'s, the vectors of the first derivative of initial ${\mathbf f}=K\bal^*$ with respect to the $\xi_j$'s. Second, besides  $\Sigma_{11}$ and $\Sigma_{01}$ terms, (\ref{e41a}) contains an extra ${\lambda_0\over{2n\lambda}}Z_0^TP\bal^*$ term, which is related to the nonlinear component $\bal^*$ projected to the perpendicular space of the $Z_1$ matrix space.

\noindent{\bf Theorem 2:}\label{t13} {\it Under the following conditions
\begin{enumerate}
\item the initial estimate $\tilde\bal$ is $\delta$ consistent, i.e. $|\tilde\bal-\bal^*|_\infty=O_p(\delta)$ for some $\delta\rightarrow0$, and
\item (\ref{e41a})-(\ref{e41c}),
\end{enumerate}
there exits some $\lambda$ with $n\lambda^2\rightarrow\infty$ such that for some constant $\eta_1>0$,  with probability $1-\exp(-\eta_1n\lambda^2)\rightarrow1$ we have results (a)-(b):
\begin{enumerate}[(a)]
\item $\hat{\ma}\subseteq\ma$ and the upper bound of $\|\hat\bxi_1-\bxi^*_1\|_\infty$ converges to
\[\label{e42}
\rho(\lambda)=\lambda\left[{{4\sigma}\over \sqrt{\tilde C_{min}}}+\|\tilde\Sigma_{11}^{-1}\|_\infty\cdot{\lambda_0\over{\lambda}}\left(n^{-1/2}\|\mathbf v^*\|_\infty+O_p(\delta)\right)+\|\tilde\Sigma^{-1}_{11}\|_\infty\right].
\]
\item If $\rho(\lambda)<\min_{j\in\ma}\xi^*_j$, then we have sparsistency of $\hat\bxi$, i.e. $\hat{\ma}=\ma$.
\end{enumerate}
}

This Theorem 2 generalizes Theorem 1 of Yuan and Lin (2007) on consistency of the linear nonnegative garrote for nonadditive models. Proof is similar to Wainwright (2009) and Ravikumar et al. (2009) which are based on the technique of a primal dual witness on model selection consistency. In Theorem 2, we use the assumption that $Z^T_1Z_1$ is invertible. Note that without this assumption, the solutions to $\hat\bxi_1$ and $\hat\bka_0$ are not unique.

In Theorem 2, $\lambda$ is required to be greater than $\sqrt{{\log p}\over n}\cdot C$, where $C$ is some constant determined by $\delta$, $\sigma^2$, and $\gamma$, so that $\exp(-\eta_1n\lambda^2)\rightarrow0$ as $n\lambda^2\rightarrow\infty$ and $(a)$ is satisfied. This places some limitation on $\lambda$ such that it can not be artificially small. Nevertheless, according to $(a)$ in Theorem 2, if we further have $\lambda+\lambda_0\sqrt{a/n}+O_p(\lambda_0\sqrt{a}\delta)+\sqrt{a}\lambda\rightarrow0$, then $\|\hat\bxi-\bxi^*\|\rightarrow0$ implying we can have estimation consistency as well.

\section{Simulation Results}\label{sec5}
\subsection{Comparison with Linear LASSO}\label{sec5.1}
In many cases, even though the underlying true model is nonlinear, variable selection using linear LASSO can be easily used since algorithms for linear LASSO are already available (e.g. LARS). These algorithms might work well as long as the following incoherence condition is satisfied,
\be\label{e34}
\left|X^T_0X_1(X^T_1X_1)^{-1}\sgn(\bbe^*_1)\right|\preceq\mathbf1,
\ee
where $X_0$ and $X_1$ are the matrices of irrelevant and relevant predictors, and $\bbe^*_1$ represents the vector of true nonzero $\beta_j$'s.

In this section we show a special case that using the NGK method sparsity of input variables can be recovered, while linear LASSO fails  due to unsatisfied  condition (\ref{e34}).

We use the same 3-variable setting by Zhao and Yu (2006) where they used simulation to demonstrate the incoherence condition in linear LASSO. First we generate iid random variables $\mbx_1$, $\mbx_2$, $\bel$ and $\mathbf e$ from $N(0,1)$ with sample size $n=100$. The third predictor $\mbx_3$ is generated by
\[\label{e43}
\mbx_3=a\mbx_1+b\mbx_2+c\mathbf e,
\]
where $a=2/3$, $b=2/3$ and $c=1/3$, and the response is generated by
\[\label{e44}
\mby=\beta^*_1\mbx_1+\beta^*_2\mbx_2+\bel,
\]
where $\beta^*_1=2$ and $\beta^*_2=3$. Denote $X_1=[\mbx_1, \mbx_2]$ and $X_0=[\mbx_3]$. Zhao and Yu (2006) showed that with this setting, $
\left({1\over n}X_0^TX_1\right)\left({1\over n}X_1^TX_1\right)^{-1}=\left({2\over3}, {2\over3}\right)$,
thus the incoherence condition (\ref{e34}) for linear LASSO is never satisfied with $\sgn(\beta^*_1)=\sgn(\beta^*_2)$.

However the incoherence condition (\ref{e41a}) of NGK provides a different incoherence condition that is satisfied. To demonstrate this, we consider using a linear polynomial kernel. Thus with $\bxi^*_1=(\xi^*_1,\xi^*_2)^T$ and $\xi^*_3=0$, we have $K(\bxi^*)=\xi^*_1\mbx_1\mbx_1^T+\xi^*_2\mbx_2\mbx_2^T$. Using the notation in (\ref{e18})-(\ref{e20}), we obtain
\be\label{e46}
\begin{split}
\tilde\Sigma_{01}\tilde\Sigma_{11}^{-1}\mathbf 1&=\left[\begin{array}{cc}a\tilde\bal^T\mbx_3\mbx_1^T\tilde\bal& b\tilde\bal^T\mbx_3\mbx_2^T\tilde\bal\end{array}\right]
\begin{bmatrix} \tilde\bal^T\mbx_1\mbx_1^T\tilde\bal  & 0                                       \\
                0                                         & \tilde\bal^T\mbx_2\mbx_2^T\tilde\bal
                \end{bmatrix}^{-1}\left[ \begin{array}{c} 1  \\  1\end{array} \right]\\
                &=a{{\tilde\bal^T\mbx_3}\over{\tilde\bal^T\mbx_1}}+b{{\tilde\bal^T\mbx_3}\over{\tilde\bal^T\mbx_2}}
\end{split}
\ee
and
\be\label{e47}
\begin{split}
{\lambda_0\over{2n\lambda}}\tilde Z_0^T\tilde P\tilde\bal
&={\lambda_0\over{2n\lambda}}\tilde Z_0^T(I-\tilde Z_1(\tilde Z_1^T\tilde Z_1)^{-1}\tilde Z_1^T)\tilde\bal\\
&={\lambda_0\over{2n\lambda}}\tilde Z_0^T\left[I-\left({1\over n}\mbx_1\mbx_1^T+{1\over n}\mbx_2\mbx_2^T\right)\right]\tilde\bal\\
&={{\lambda_0\over{2n\lambda}}}\tilde\bal\left(\mbx_3\mbx_3^T-a\mbx_3\mbx_1^T-b\mbx_3\mbx_2^T\right)\tilde\bal.
\end{split}
\ee
In equations (\ref{e46})-(\ref{e47}), we use the fact that for independent random normals, ${1\over n}\mbx_i^T\mbx_j=\delta_{ij}, i, j=1,2$ and ${1\over n}\mbx_3^T\mbx_j=a \hbox{ or } b$ for $j=1 \hbox{ or } 2$. Given $\tilde\bal=(\lambda_0I+K(\tilde\bxi))^{-1}\mby$ with $\tilde\bxi=(1,1,1)^T$, we can calculate the left hand side of (\ref{e41a}). For demonstration with one simulation example, we calculate two incoherence condition curves vs $\lambda$ and $\lambda_0$, respectively. For the first curve vs. $\lambda$, we fix $\lambda_0=0.0026$ estimated by REML. For the second curve, we fix $\lambda=1.516$, where we choose the model with minimum BIC and vary $\lambda_0$.

\begin{figure}[bth]
\begin{center}
\includegraphics[height=80mm]{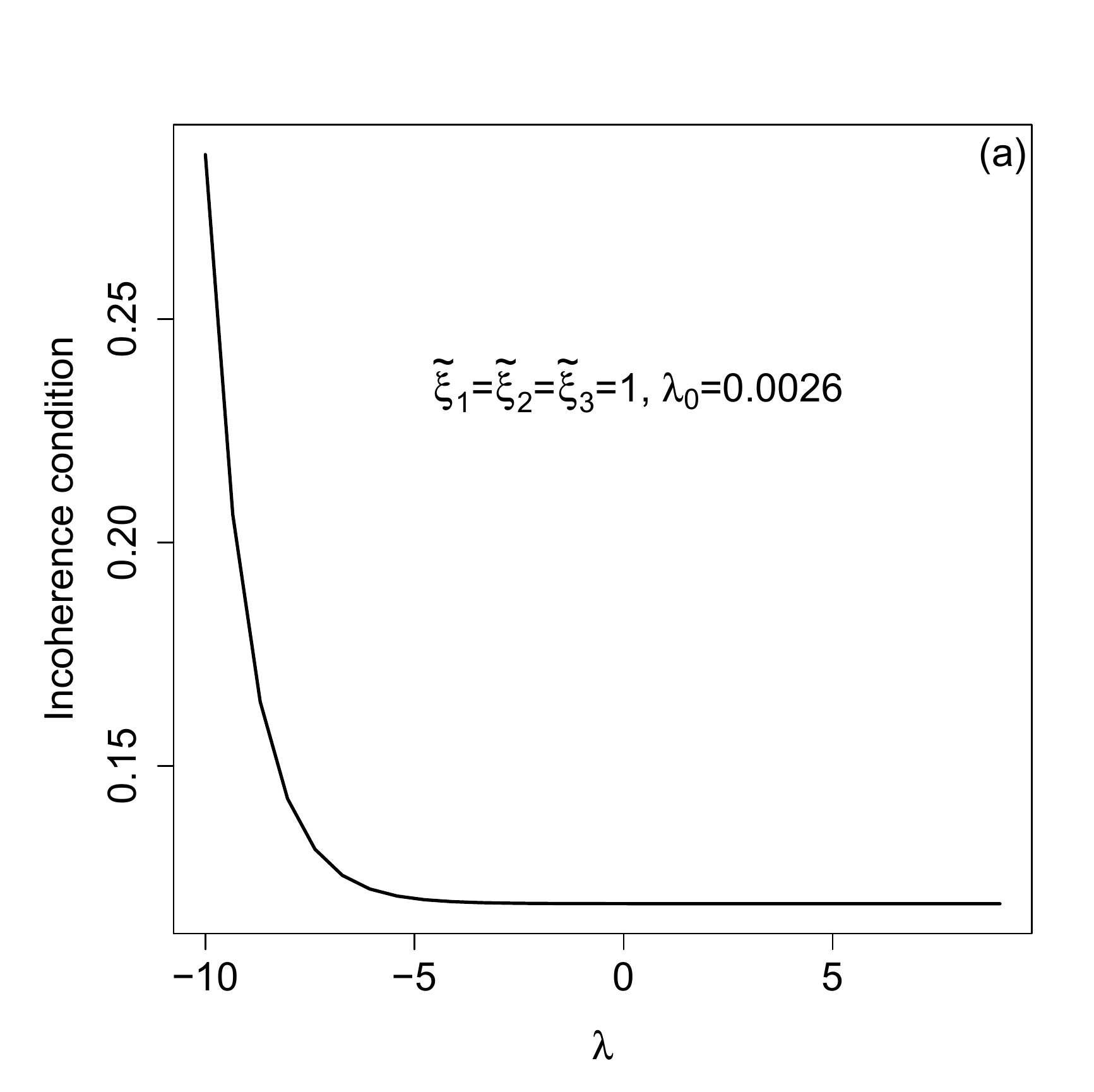}\includegraphics[height=80mm]{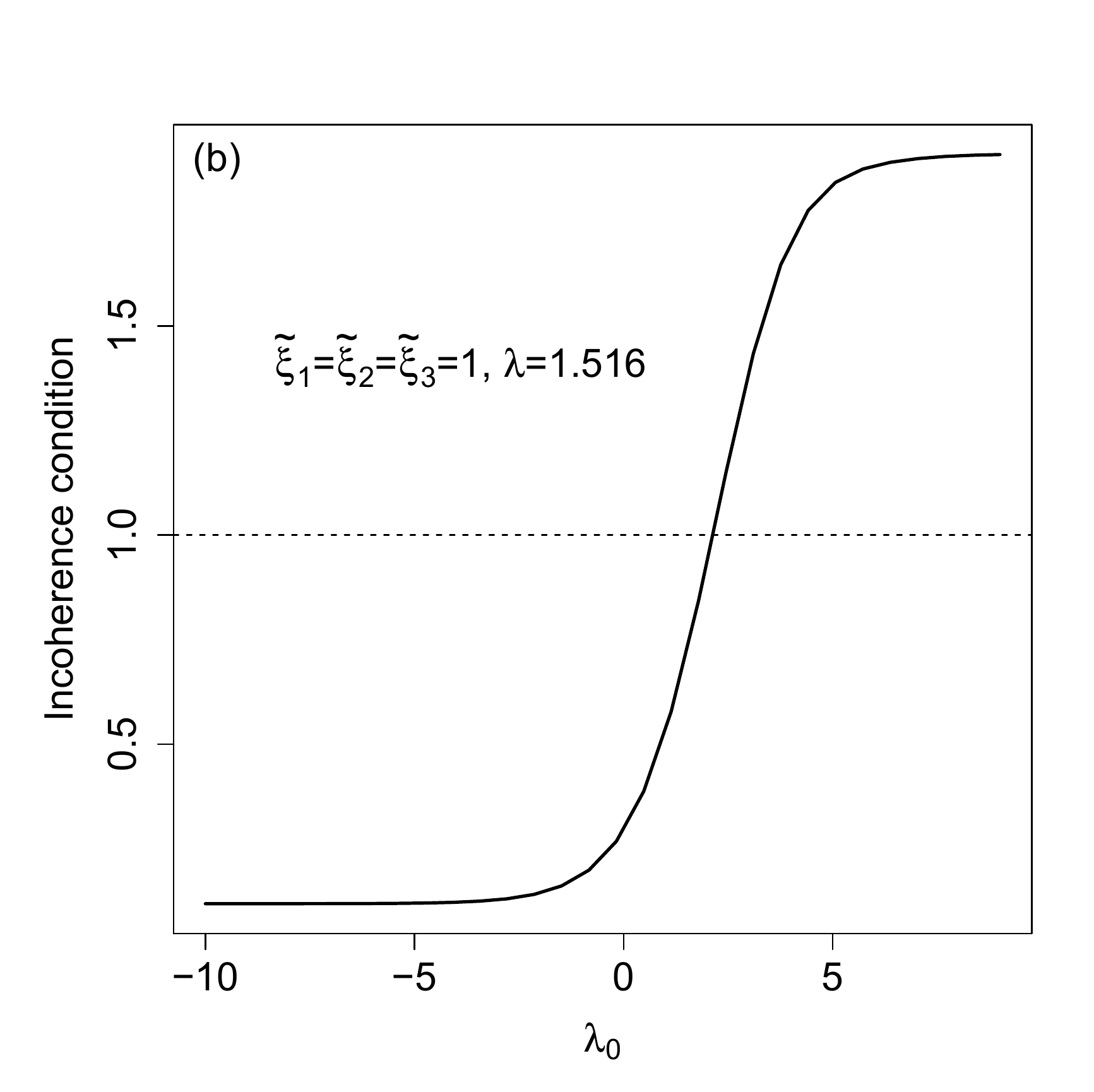}
\includegraphics[height=80mm]{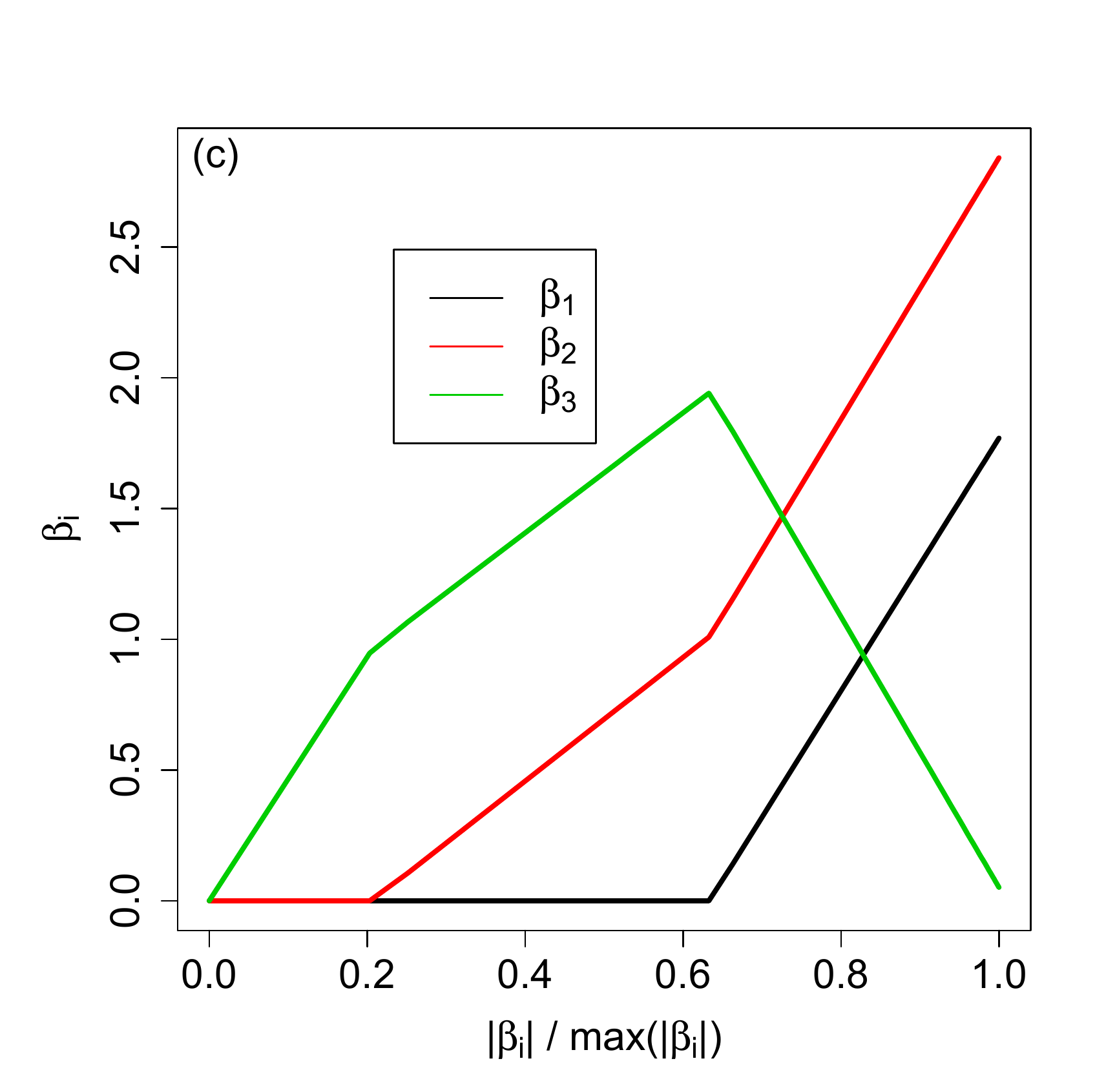}\includegraphics[height=80mm]{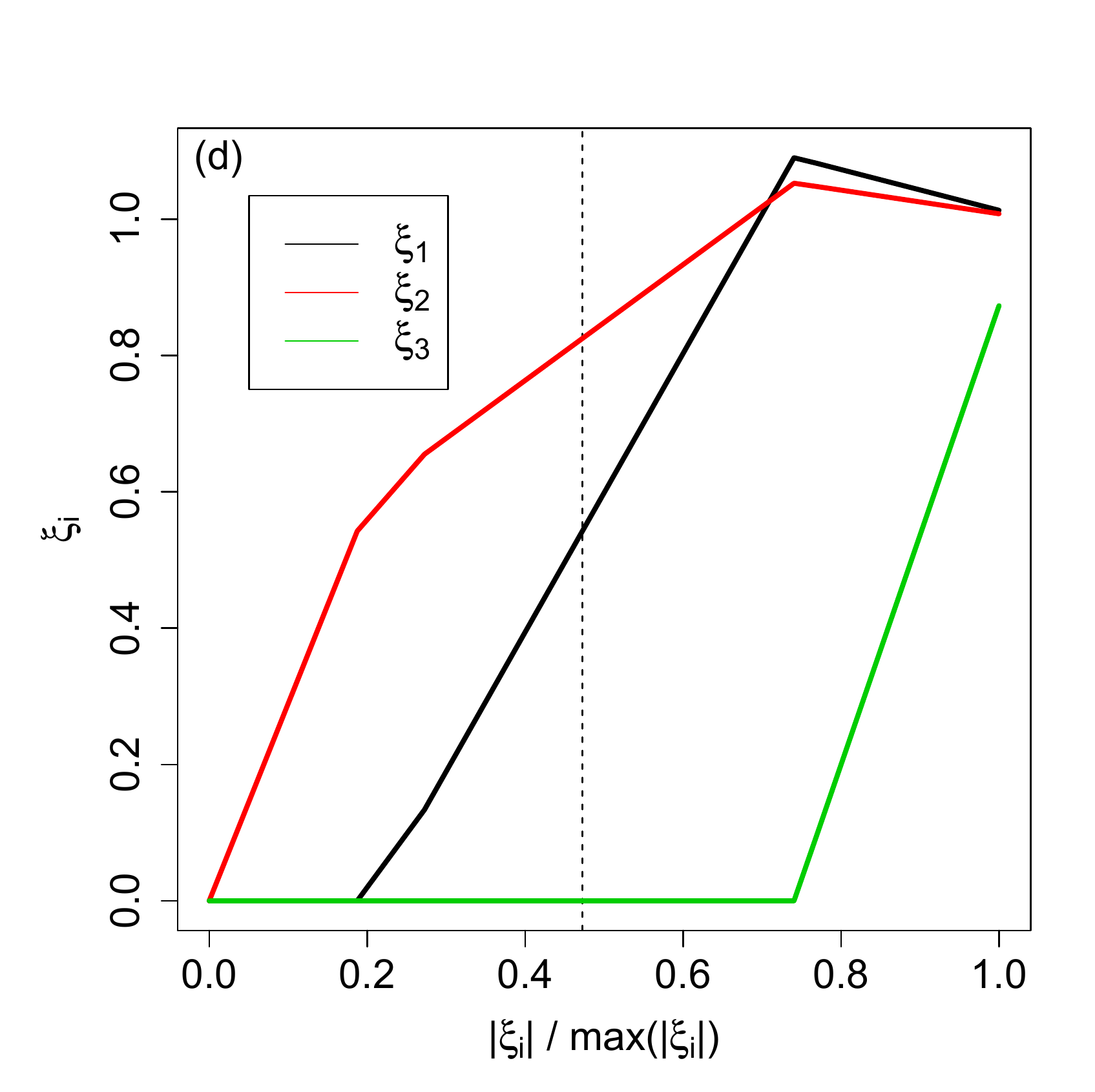}
\caption{ (a) Incoherence condition values vs. $\lambda$ with $\lambda_0$ fixed at 0.0026, (b) Incoherence condition values vs. $\lambda_0$ with $\lambda$ fixed at 1.516, (c) solution path of $\beta_i$'s for linear LASSO, and (d) solution path of $\xi_i$'s for NGK. All plots use initial $\tilde\bal=\Delta^{-1}(\tilde\bxi)$ with $\tilde\bxi=(1,1,1)^T$.}\label{f1}
\end{center}
\end{figure}

Figure \ref{f1}(a)-(b) show two plots: one (a) is for the incoherence condition values vs. $\lambda$ and the other (b) is for the incoherence condition values vs. $\lambda_0$. They show that for certain $\lambda$ and $\lambda_0$ values, the incoherence condition values are smaller than one, thus condition (\ref{e41a}) is satisfied and there is a possibility that the variable selection procedure of NGK can recover sparsity of those irrelevant variables.

Figure \ref{f1}(c)-(d) also show plot of the regularization path for linear LASSO (c) and NGK (d). It can be seen that for linear LASSO, $\beta_3$ is always non-zero on the path except when $\lambda=0$, which means linear LASSO will always select $\beta_3$. However, the regularization path of NGK shows that for some $\lambda$, $\xi_3=0$, but both $\xi_1$ and $\xi_2$ are greater than zero, providing the possibility to select the correct variable set. The dashed line in Figure \ref{f1}(d) indicates where we based model selection on minimum BIC.

\subsection{Simulation Example 1}\label{sec5.2}
In this section we test the implementation of NGK on a nonlinear multiple regression simulation scenario. For this scenario, we consider the simple situation where the total number of predictors $p$ is $11$ and the first $a=5$ predictors are relevant. Three settings with sample sizes $n=64, 128$ and $256$ are compared. For each setting, a total of 200 runs were generated.

The nonlinear function $\mathbf f$ was generated using a stationary zero mean Gaussian process
\[\label{e48}
\mathbf f\sim N\left(\mathbf 0, \sigma^2_\alpha K(\bxi^*,X)\right),
\]
where we use Gaussian kernel $K(\bxi^*,X)$ with $X=[\mbx_1,...,\mbx_p]$, with each column generated independently by $\mbx_j\sim U(-2.5, 2.5)$ , and $\xi^*_1=...=\xi^*_a=2$. The responses were then produced by
\be\label{e49}
\mby=\mathbf f+\bel,
\ee
where $\bel\sim N(\mathbf 0, \sigma^2 I)$. In this scenario, we chose $\sigma^2_\alpha=10$ and $\sigma^2=1$ ($\lambda_0=1/10$) to produce our datasets. Note that model (\ref{e49}) is equivalent to $\mby=K\bal+\bel$ with $\bal\sim N\left(\mathbf 0, \sigma^2_\alpha K^{-1}\right)$.

We also note that, in this example, $f$ is not a fixed function anymore, and $\bal$ is random. This should lead to different incoherence conditions and, because of the randomness of $\bal$, the probability of recovering sparsity is expected to be lower. However, we only use this example to demonstrate the performance of NGK with a similar variable selection method like COSSO.

The solution paths and BIC curves of one simulation run for NGK with a Gaussian or linear polynomial kernel are shown in Figure 1 of Supplementary Materials. The frequency of variables selected in the model for 200 runs are also summarized in Table 1 of Supplementary Materials.

Five statistics are reported in Table \ref{t2}. They are ``False Positive Rate (FP-rate)'', ``False Negative Rate (FN-rate)'', ``Model Size (MS)'', ``Residual Sum of Squares (RSS)'' and ``Squared Error (SE)'', where $\hbox{FP-rate}={\#{False\,Positive}\over{\#False\,Positive + \#True\,Negative}}$,\\ $\hbox{FN-rate}={{\#False\,Negative}\over{\#False\,Negative + \#True\,Positive}}$, $\hbox{RSS}=\sum_i^n(y_i-\hat f_i)^2/n$ and $\hbox{SE}=\sum_i^n(f_i-\hat f_i)^2/n$ are calculated for each individual run. The average and standard deviation of these statistics from 200 runs are reported. $\hat{\mathbf f}$'s are calculated using  least squared error estimation of the kernel machine with corresponding $\hat\bxi$, i.e. $\hat{\mathbf f}=K(\hat\bxi)\Delta^{-1}(\hat\bxi)\mby$. SE can be used to assess the accuracy of estimation of the nonlinear function $f$. Note that in Table \ref{t2} and the following tables and plots, ``NGK Gauss'' and ``NGK Poly'' represent NGK method with Gaussian kernel and with linear polynomial kernel, respectively.

The performance of COSSO and NGK methods are similar. COSSO is slightly better in terms of the FP rate and FN rate. However, we consider the linear polynomial kernel NGK method as the best method in this example, not only because it has similar FP and FN rates as COSSO but also because it shows the best accuracy in estimating $f$. In addition, the Gaussian kernel NGK method also has higher accuracy in estimation than COSSO. As expected, we can see all methods have comparable high FN rates since the function $f$ is not fixed.

\subsection{Simulation Example 2}\label{sec5.3}
In this example, we consider fixed $f$ and generate response $y$  using
\[\label{e50}
y=f+\epsilon=10\cos(x_1)+3x_2^2+5\sin(x_3)+6\exp(x_4/3)x_4+8\cos(x_5)+x_5x_2x_1+\epsilon,
\]
where $\epsilon\sim N(0,1)$ and $x_j\sim U(0,1), j=1,...,p$. Function $f$ in this simulation is similar to the one used in Liu et al. (2007). In this example, we consider $o=10$ total predictors where the first $a=5$ are relevant. Again, three settings with sample sizes $n=64$, $128$, and $256$ were generated with a total of 200 runs per setting.

A selected example of the solution paths for two NGK methods and the BIC curves are shown in Figure 2 of Supplementary Materials. The selection frequency of 200 runs are listed in Table 2 of Supplementary Materials. Five statistics of 200 runs are summarized in Table \ref{t4}. It can be seen that all three methods have the same zero FN rate. When the sample size is $n=64$, the COSSO approach seems to perform the best in terms of the FP rate, but still has the worst estimation accuracy. When sample size increases, NGK methods quickly catch up in terms of FP rate. When $n=256$, the three methods have nearly the same FP rate. It can be seen that with increasing sample size, COSSO maintains almost the same estimation accuracy, while NGK methods seem to estimate more accurately. In this example, the Gaussian kernel NGK method is considered to be the best method not only because it performs as well as the other methods in terms of FN and FP rates, but also because it has the best estimation accuracy.

According to Example 1 and Example 2, we observe that although the COSSO method is based on only an additive model (or at most second order interactions), it is capable of variable selection for models with higher order interactions. However, when higher order interactions are included in the true model, additive type methods may not perform as well as the kernel based methods in terms of estimation accuracy. In Example 1, the interactions of the model might be any order since we use a Gaussian process to generate the data, and in Example 2, the true model contains third order interactions, but the COSSO procedure we apply only models the additive main effects. In contrast to this, instead of modeling each interaction component, the Gaussian kernel NGK method can model interactions of any order, as well as select the input variables correctly.

\subsection{Simulation Example 3}\label{sec5.4}
As mentioned before, when $p>n$, COSSO and LARS on nonnegative garrotes both fail. In this example, we consider a special case with $p=80$ and $n=64$. So far there is no other approach capable of modeling the nonadditive model and selecting predictors for $p>n$ cases. Hence we only compare the Gaussian and linear polynomial kernel NGK methods using our backfitting algorithm. Example 3 has the same true function as Example 2. The first five predictors are relevant and a total of 400 runs have been simulated. Since computing becomes intensive when $p$ is large, we only demonstrate the results with $n=64$.

\begin{figure}[bth]
\begin{center}
\includegraphics[height=80mm]{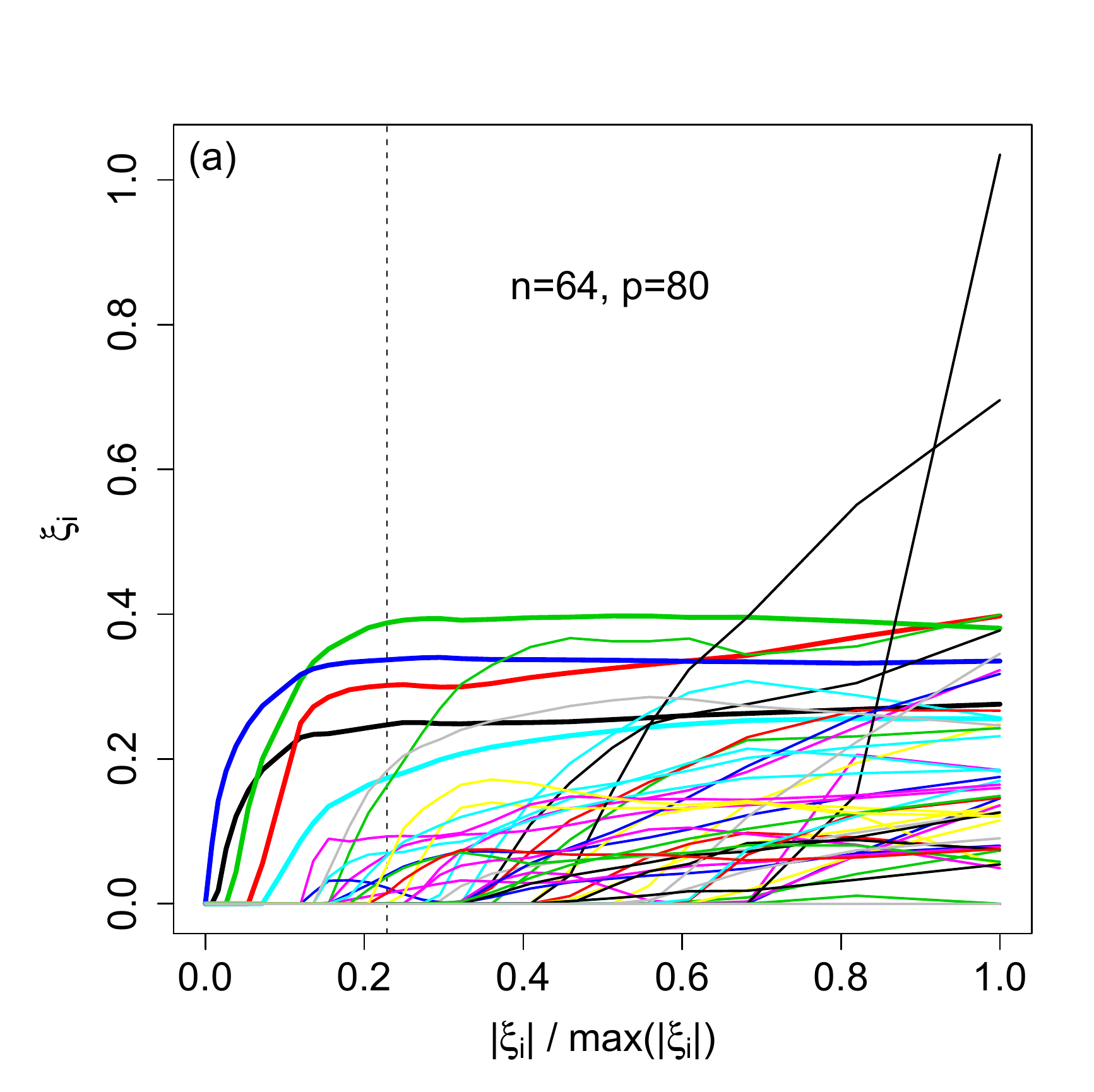}\includegraphics[height=80mm]{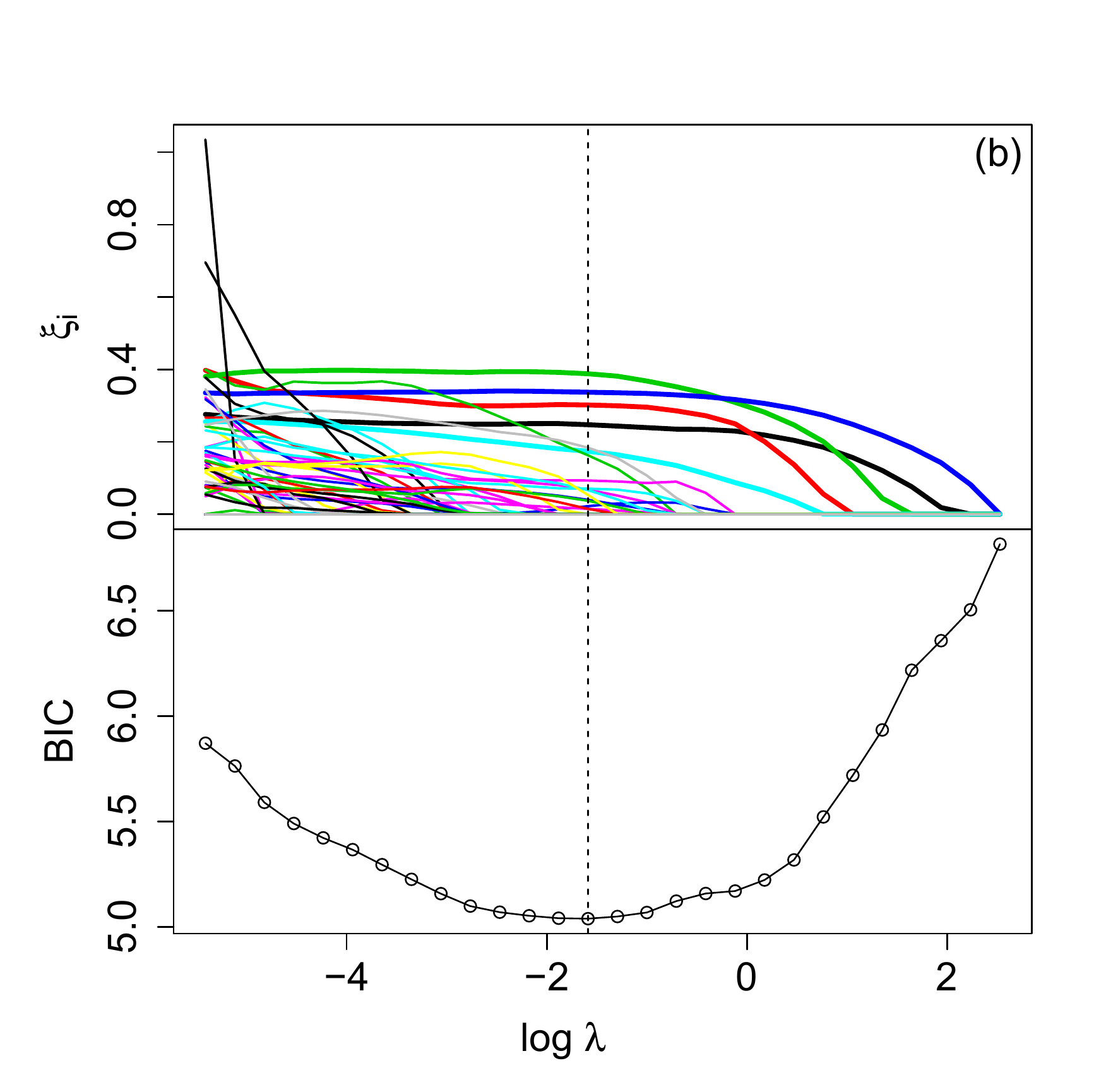}
\includegraphics[height=80mm]{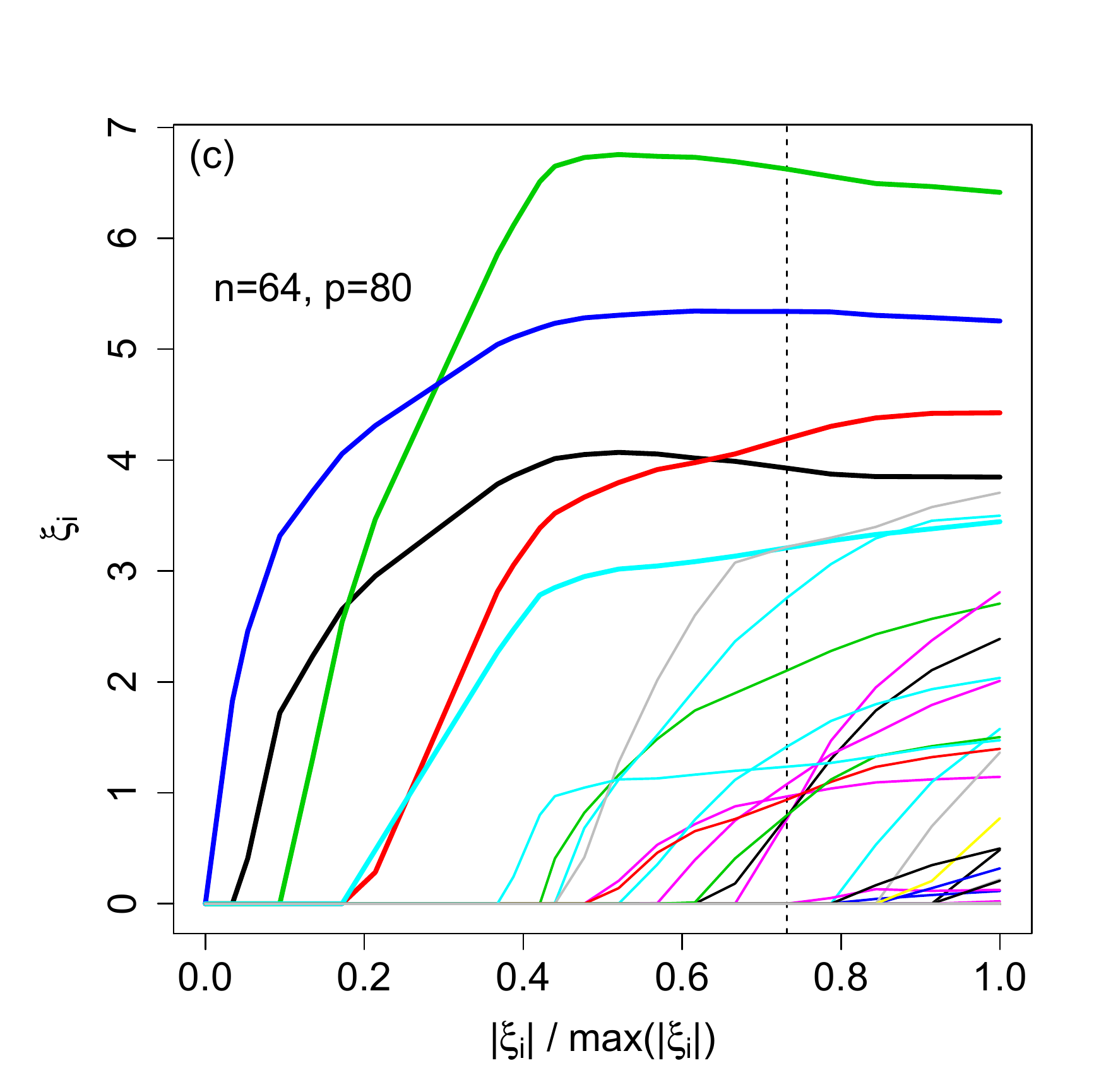}\includegraphics[height=80mm]{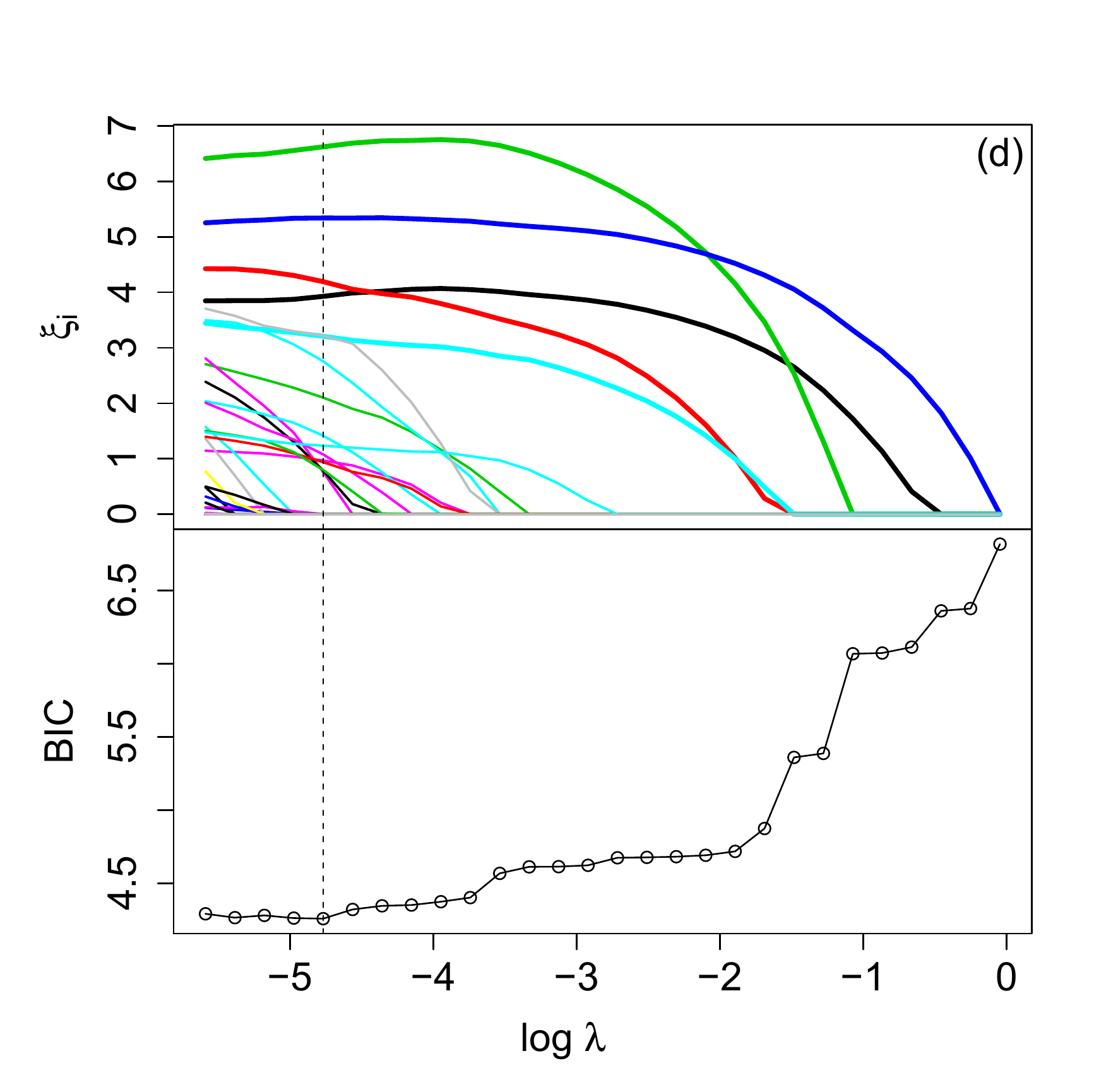}
\caption{Selected example of NGK solution path for Simulation Example 3 using Gaussian kernel, (a) and (b), and linear polynomial kernel, (c) and (d). Left side:  $\xi_j$'s vs. $L_1$ norm of $\xi_j$'s, Right side:  $\xi_j$'s and BIC vs. $\log\lambda$.}\label{f5}
\end{center}
\end{figure}

Figure \ref{f5} shows example solution paths for Example 3 by the Gauss and linear polynomial kernel NGK methods. In both cases the number of variables selected by BIC is greater than 5. Other criteria such as Cp and CV in this simulation also select larger model sizes. Therefore, in Example 3, variable selection according to  a single run is not sufficient for revealing the correct model.

\begin{figure}[bth]
\begin{center}
\includegraphics[height=110mm]{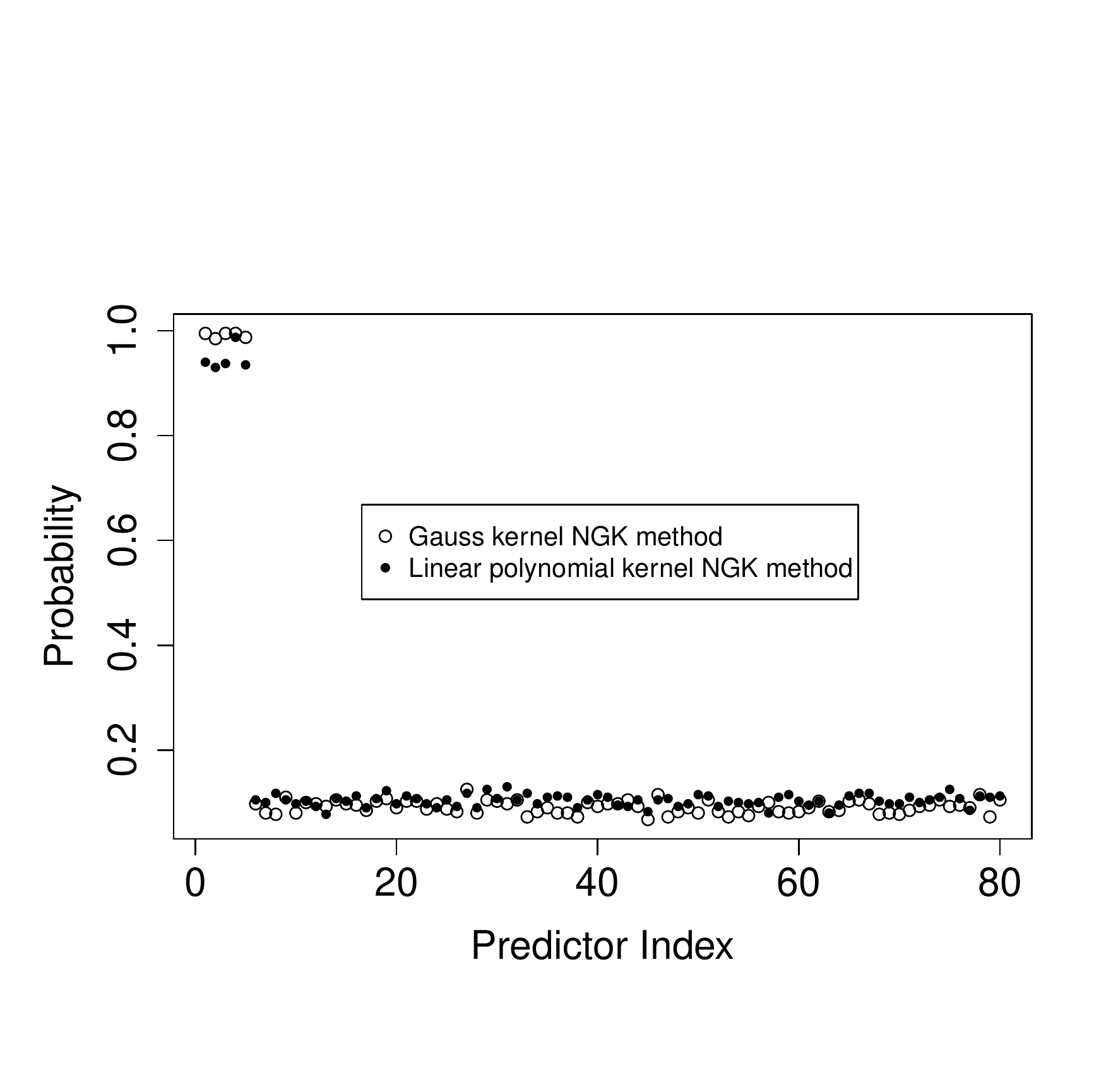}
\caption{Selection probability of each predictor in Simulation Example 3 for 400 runs using two NGK methods.}\label{f6}
\end{center}
\end{figure}

Because of the number of predictors, instead of a table, we portray the selection frequency or probability of each variable for 400 runs in Figure \ref{f6}. There is little difference between the two NGK methods in terms of the selection probability. This can be seen from Figure \ref{f6} where the first five variables have selection probability very close to 1.0 for both methods. In addition both methods show the same behavior in that the first five variables are clearly separated from the remaining 75 variables in terms of selection probability. However, the linear polynomial kernel method has a slightly higher FN rate than the Gaussian kernel approach since these five probabilities are slightly higher for the Gaussian kernel method.

From Table \ref{t5} we can see the FP and FN rates for Example 3. Compared to Example 2, the FP rate of Example 3 for the Gaussian kernel approach is comparable, 0.09 and 0.08, respectively. For the linear polynomial kernel method, the FP rate increases slightly. The major difference is that in Example 2 FN-rates  are zero for both methods, but are nonzero in Example 3. This is reasonable since inclusion of many irrelevant predictors deteriorates variable selection performance. One additional observation from example 3 is that the standard deviation of the five statistics across 400 runs is much larger for the polynomial kernel method while the average is similar between both methods. This is because we base selection on BIC. While the BIC curve for the Gaussian kernel method has a clear minimum, the BIC curve for the linear polynomial kernel method drops from $\xi_j=0$ and becomes flat. We select the model at the turning point of the BIC curve which may introduce some variability for different runs. We also realized the the average model size is greater than 5  from Table \ref{t5}, which reflects the fact that the including more irrelevant predictors will result in more irrelevant predictors being selected.

The above simulation results suggest that if we choose the model according to BIC or any other criteria based on only one run, we may select more irrelevant predictors. However, if we select the model based on the frequency or probability of selected predictors, as in Figure \ref{f6}, it is clearer that the five true variables behave differently from the others. This provides new ideas regarding variable selection: it is less powerful to select the correct model with one single set of data than with multiple drawing of the data. Furthermore, if we use multiple drawing or resampling of the observations, we can estimate the selection probability of each variable which provides more power to select the correct variables. In the following section we apply this idea to two real data sets.

\section{Applications}\label{sec6}
In this section, we describe the application of our method in two practical settings.
\subsection{Key Selection For Cryptography Data}\label{sec6.1}
Our first example is taken from a cryptography study. Side-channel analysis (SCA) is
a technique of cryptanalysis with which an attacker estimates the secret key based on
information gained from the physical implementation of a cryptographic algorithm. Figure 3(a) in Supplementary Materials gives the diagram of attacking on an Advanced Encryption Standard (AES) system. The outside box represents an electronic circuit system that implements the AES
algorithm. The AES algorithm processes the input data ``$in_k$'' and produces an encrypted output ``$out_k$'' using a secret key with bytes $\theta_k, k=0,...,15$. The 16 S boxes each takes an 8-bit byte $\theta_k$. The attacker's objective is to determine the value of the secret key $\theta_k$'s.
The output of the AES algorithm is captured in each of the $16$ encryption rounds, and the corresponding power consumption of all rounds is recorded as $\mby$. By observing the output, one can therefore infer 16 estimates $X_k=[\mbx_{1,k},\mbx_{2,k},\cdots,\mbx_{256,k}], k=0,...,15$, corresponding to 16 secret key bytes, and there are $2^8=256$ possibilities for each estimate with only one as the true.  The SCA proceeds by observing $n$ encryptions. The data structure can be expressed in a matrix-form as shown in Figure 3(b) of Supplementary Materials, where $\mby$ is an $n\times1$ vector, and each $X_k$ is an $n\times 256$ matrix. Note that each $x_{k,j}, j=1,...,256$, is a function of the output $out_k$ and the $j$th key guess ${\hat\theta_{k,j}}$ on the $k$th S box. The SCA problem is to find the set of columns that represents the true $\theta_k$'s. The index of the selected column in each $X_k$ returns the value of secret key byte $\theta_k$.

Define $X=[{X_0,\ldots,X_{15}}]$ to be an $n \times(r\times k)$ matrix, reflecting internal estimates for a SCA with $n$ measurements, $k$ key parts of $r$ guesses per part.  In the SCA example, $n=5120$, $r=256$ and $k=16$. The objective is to identify which possible key guesses (or what combination of the columns of $X$) are highly associated with the power consumption trace $\mby$. Since there are $k$ key bytes and $r$ possible key guesses for each key byte, there are a total of $(r\times k)\choose k$ possible ways to select $k$ variables. The power consumption trace $\mby$ can be expressed in terms of the key estimates $X$ by model (\ref{e1}). It is reasonable to assume that there are no interactions among the key guesses, so that we can use a linear polynomial kernel NGK method for these data.

The data set $(\mby,X)$ contains 5120 observations and a total of $p=16\times256=4096$ predictors. Identifying the 16 key bytes is a variable selection problem. Due to the high dimensionality of the $X$ space, directly applying our NGK approach is less efficient. Fan and Lv (2008) discussed sure independence screening (SIS) for ultrahigh dimensional feature spaces, and Fan et al. (2011) extended correlation learning in linear models to nonparametric independence screening (NIS) in additive models. They argue that under certain conditions, the probability of the screened model including all the relevant predictors, approaches one as $n$ increases. We adopt a similar procedure, NIS-NGK, that is, we apply NIS and then perform the NGK variable selection approach using a polynomial kernel.

Another issue for this dataset is the observation size. With $n=5120$, Computation becomes expensive due to calculation of the kernel $K$, especially when using a Gaussian kernel. We take a resampling approach with observation size $m=2048$ to reduce the computation burden. However, it turns out that variable selection on large datasets through multiple resampling is more powerful in identifying the significant predictors than selecting the predictors based on a single run. There is not much work discussing the resampling/bootstrapping procedure in variable selection. Hall et al. (2009) proposed a m-out-of-n bootstrap on linear LASSO and provided theoretical justification of their resampling approach. We extended this m-out-of-n bootstrap approach to our NGK method.

The screening approach we applied is meant to rank the predictors according to the descent order of the residual sum of squares by a marginal nonparametric regression.
\be\label{e51}
\mathcal{S}=\{1\le j\le p: r_j\le C\},
\ee
where $r_j=\min_{\xi_j\bal}\|\mby-\xi_jD^j\bal\|^2$ and $C$ is a predefined threshold value depending on $n$. To reduce computational time, we take $\bal=\mby$ and all $\xi_j=1$. Then the NIS screening is equivalent to SIS which ranks the predictors by the correlation $\mbx_j\mby^T$. This can be seen by plugging $\bal=\mby$ and $\xi_j=1$ into $r_j$, $r_j=\mby^2-(\mbx_j\mby^T)^2$. Using this approach we first screen the predictor size down to 200. According to the Theorem 1 in Fan et al. (2011), with $n$ very large and finite predictor size, the probability of the screened predictors including the true predictors is close to one.

Following the NIS step, we apply the NGK variable selection approach to one resampled dataset from the original $5120$ observations. Figure 4 in Supplementary Materials is an example of the m-out-of-n resampling from the $n=5120$ observations. The solution path in Figure 4(a) in Supplementary Materials shows that there are 16 predictors (bold lines) which behaves differently from the others. Because we have information about the AES key,  we know there are 16 bytes corresponding to 16 true predictors. By checking the AES key, these 16 predictors are the exact 16 key bytes. Additionally, the BIC curve shows that the 16 predictors are clearly separated from the rest of the points on the curve (see Figure 4(b) in Supplementary Materials ). Other critera such as Cp and GCV, all show the same behavior. It is obvious that these 16 predictors should be selected for the true model. However, if we use minimum  BIC as our model selection criterion, we will have  a total of 38 predictors selected in this run. Although these 38 predictors include the 16 true predictors, we are over selecting. This is true even when we sample more observations or use all 5120 observations.

\begin{figure}[bth]
\begin{center}
\includegraphics[width=100mm]{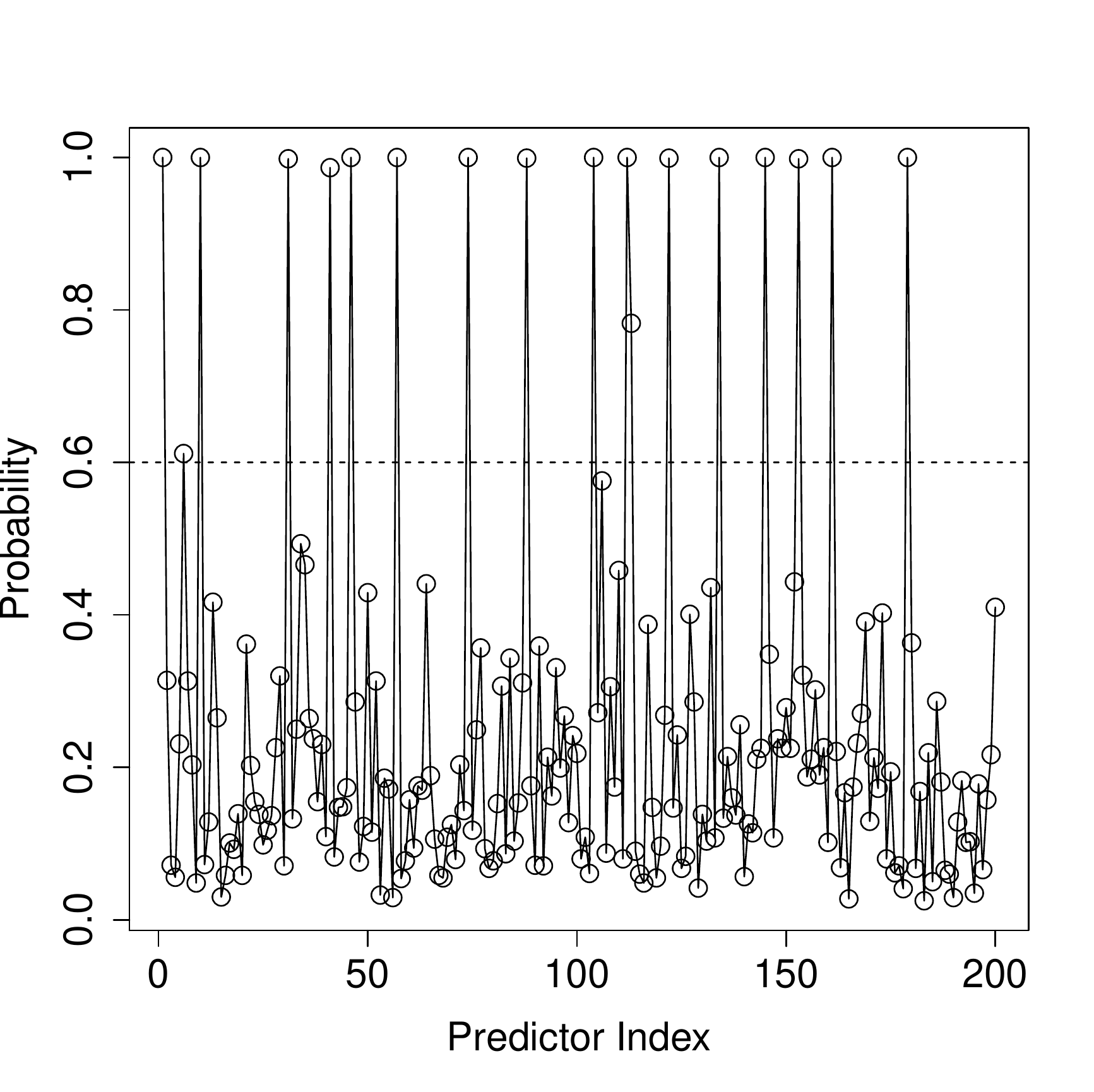}
\caption{Selection probability of each key guess of SCA data using m-out-of-n resampling procedure, $m=2048, n=5120$ and total 1200 runs.}\label{f9}
\end{center}
\end{figure}

Thus we further resample the dataset up to 1200 runs with replacement with each run choosing variables according to the BIC criterion, and we count the frequency of selected variables. Since we observed that the BIC minimum usually occurs when around 50 variables are selected, in order to reduce computation, we use a selection window such that we choose the first 50 predictors in the model for each resampling. The probability/frequency of being selected for 200 predictors are plotted in Figure \ref{f9}. Use $60\%$ as the selection threshold, we will choose 18 predictors based on Figure \ref{f9}. If we use the threshold probability $80\%$, we can exactly choose the true 16 key bytes. Through resampling from  a large dataset, we are able to simulate selection probability, and this process of variable selection is more powerful than simply relying on one fixed data set using usual criteria.

\subsection{Gene Selection in Pathway Data}\label{sec6.2}
We next apply our Gaussian kernel NGK method to a set of diabetes data from Mootha et al. (2003). They provided pathway based analysis to classify two phenotypes, 17 normal and 18 Type II diabetes patients. A pathway is a predefined set of genes that serve a particular cellular or physiological function. They showed that pathway based analysis can detect coordinate subtle changes among a set of genes. It is known that genes in a pathway are not independent of one another and interact with unknown structure. The top significant pathways related to the diabetes disease have been identified (Mootha et al., 2003). Pathway 133 (``Oxidative phosphorylation''), pathway 4 (``Alanine and aspartate metabolism'') and pathway 140 (``MAP00252-Alanine-and-aspartate metabolism'') are three interesting ones which contain a total of 58, 18 and 22 genes, respectively.

For each pathway we label the genes by their appearance index, gene $1$, gene $2$ ... and so on. Note the same gene index from two pathways does not imply the same gene. Since the 18 genes in pathway 4 are all included the 22 genes in pathway 140, we use the gene index of pathway 140 to label genes in both pathways. Thus gene 4, 5, 19 and 20 do not appear in pathway 4. Hence, in this application, the data set structure is $(\mby, X)$ with a total of $n=35$ observations and $p=58, 18$ and $22$ predictors, respectively. The response is the outcome of glucose levels.

\begin{figure}[bth]
\begin{center}
\includegraphics[height=72mm]{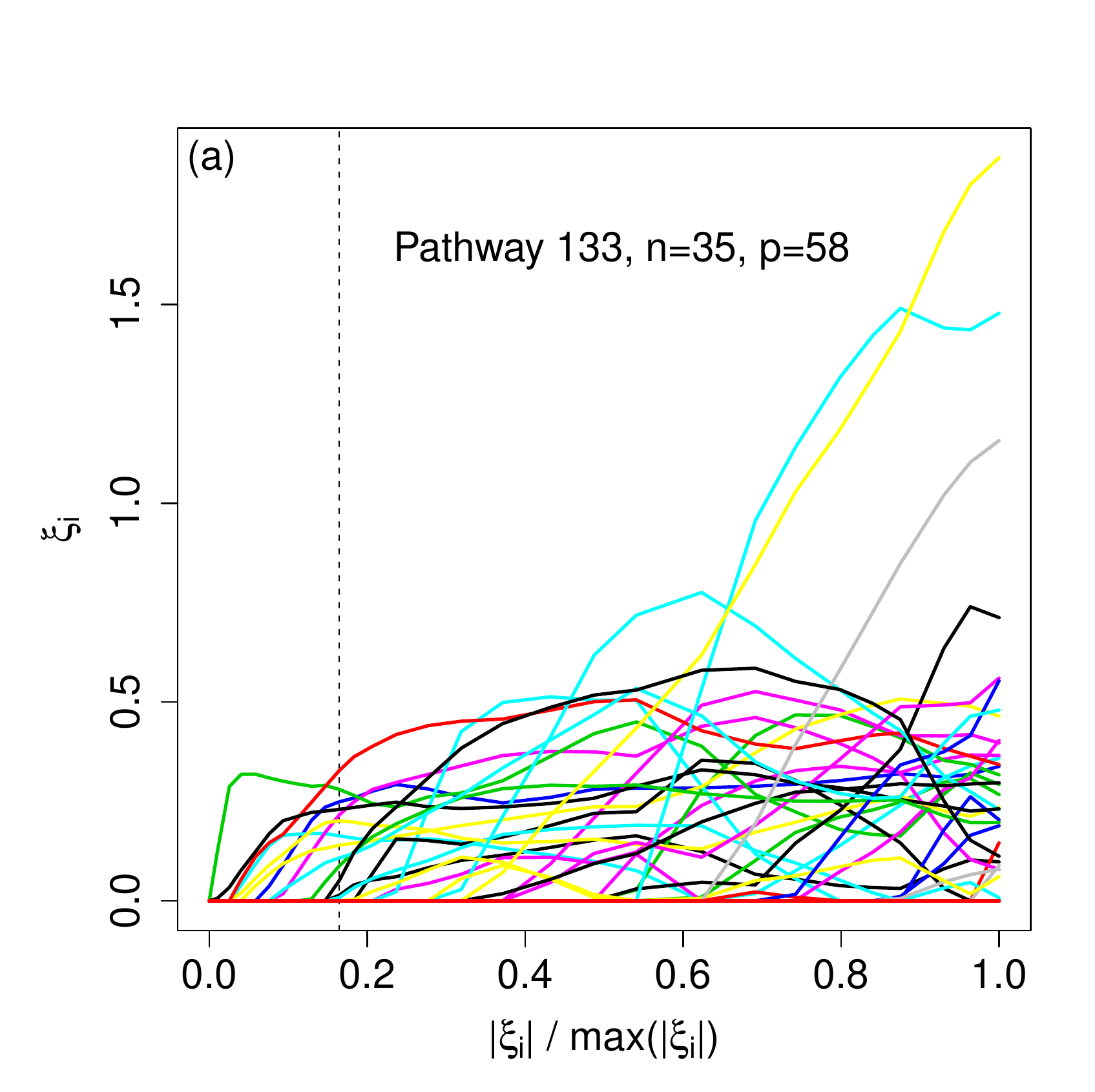}\includegraphics[height=72mm]{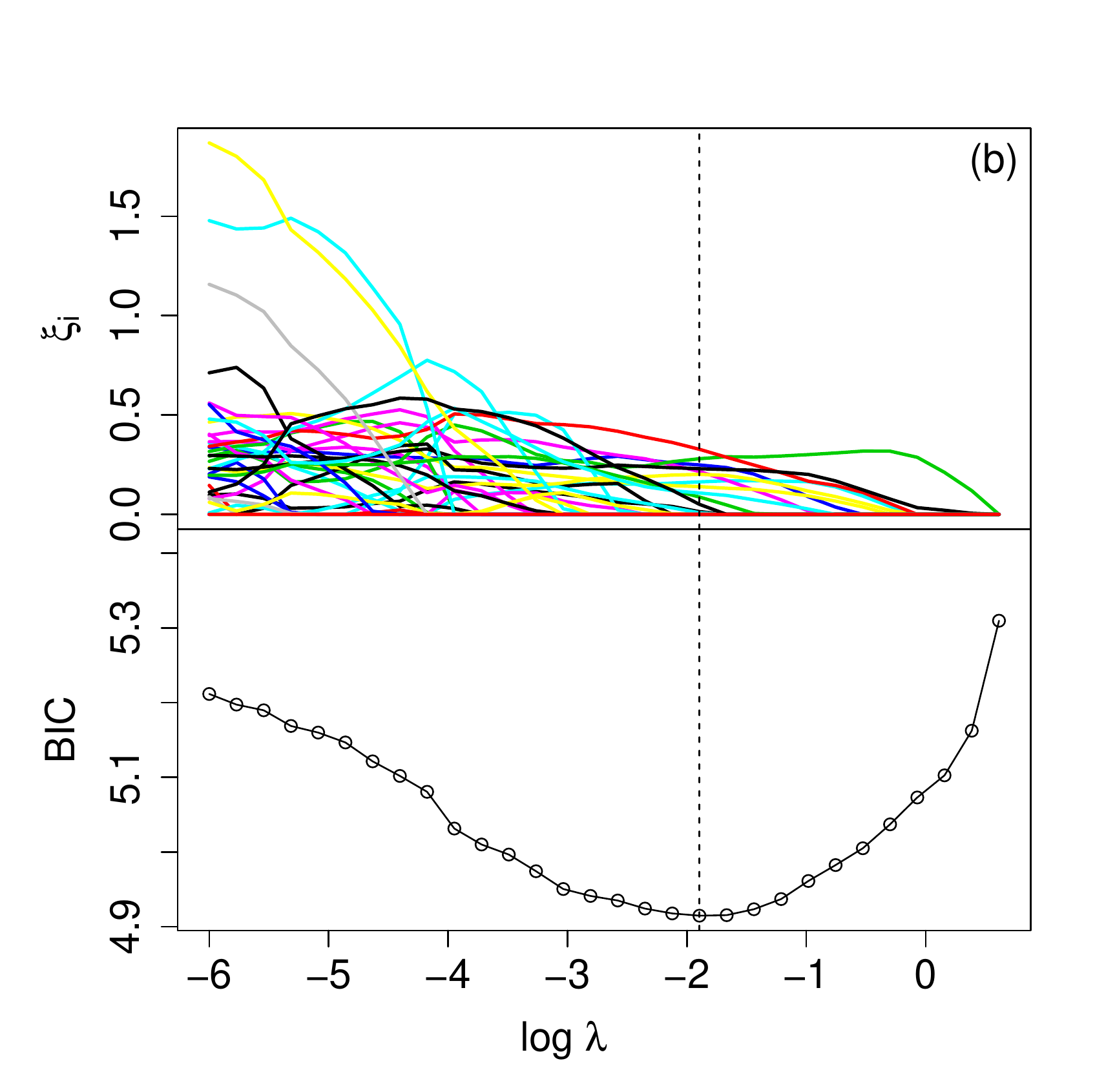}
\includegraphics[height=72mm]{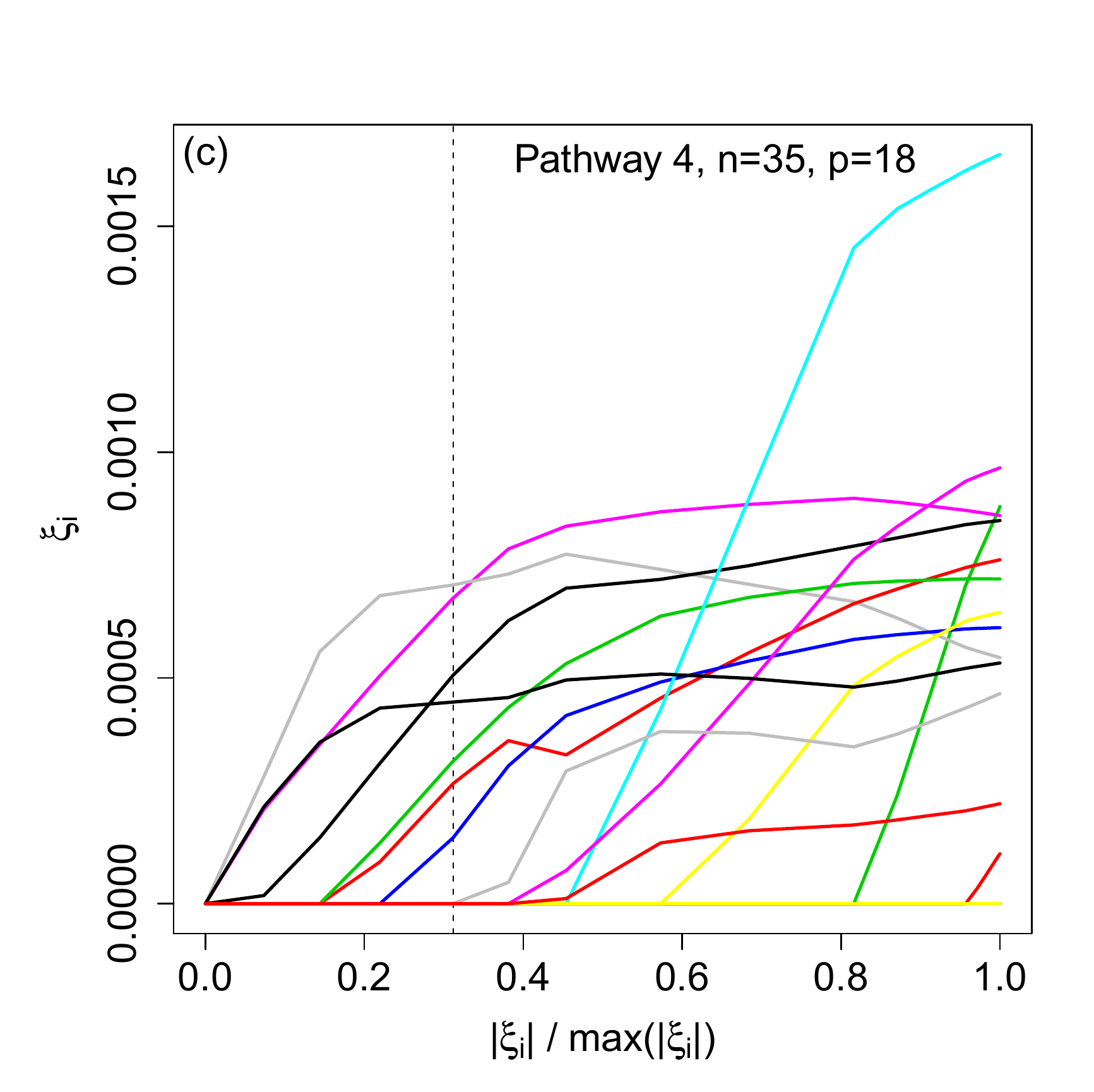}\includegraphics[height=72mm]{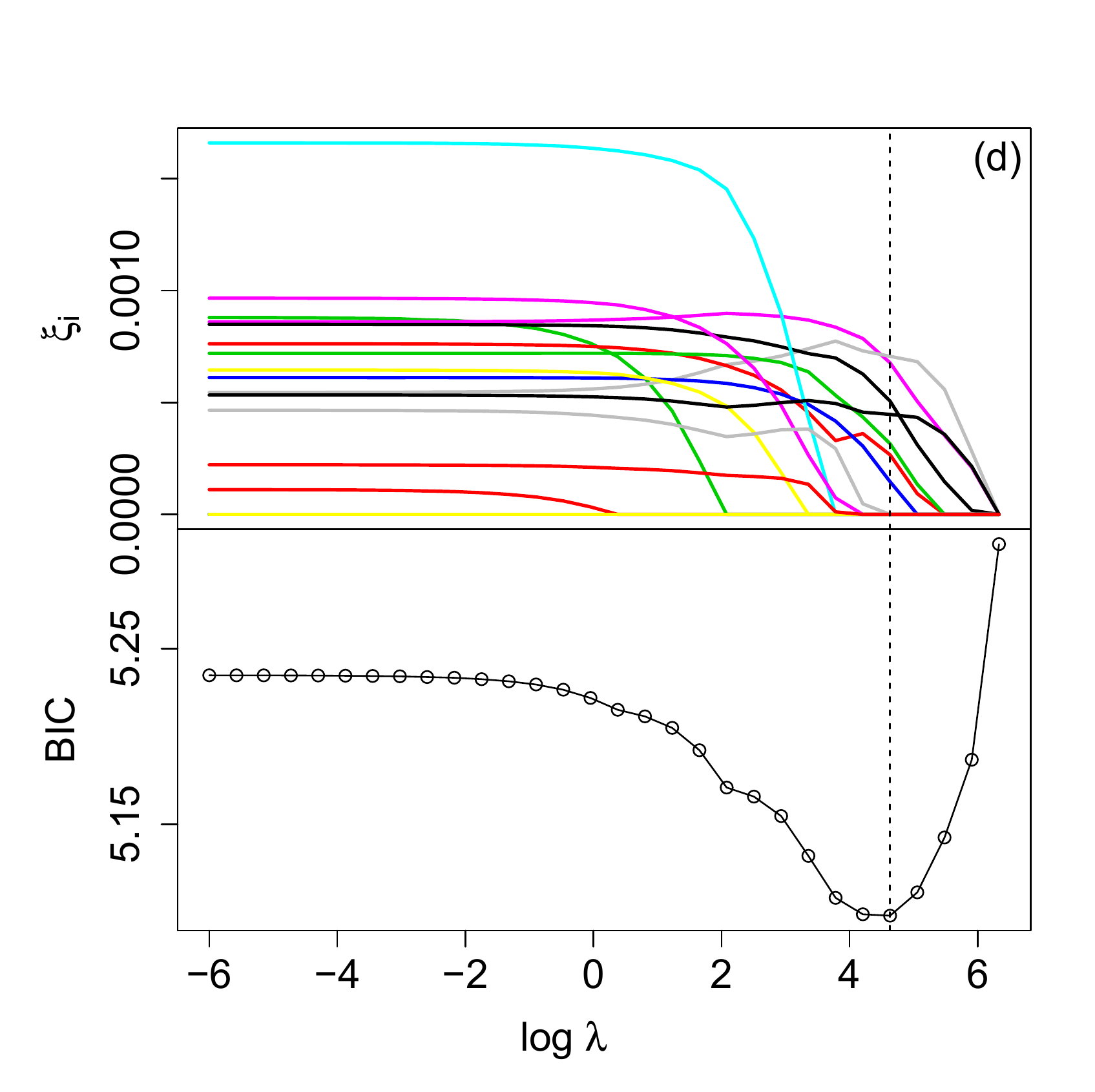}
\includegraphics[height=72mm]{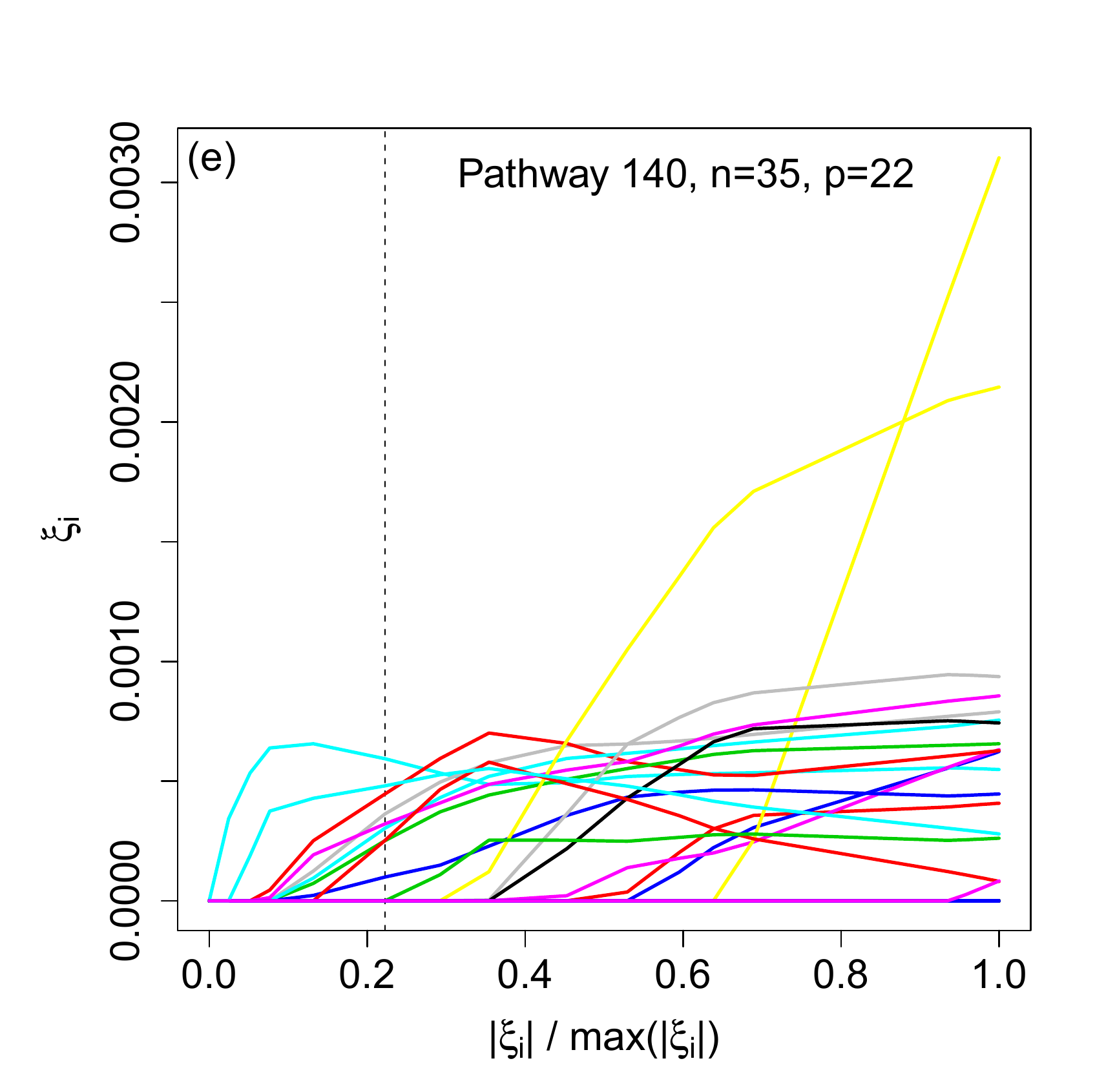}\includegraphics[height=72mm]{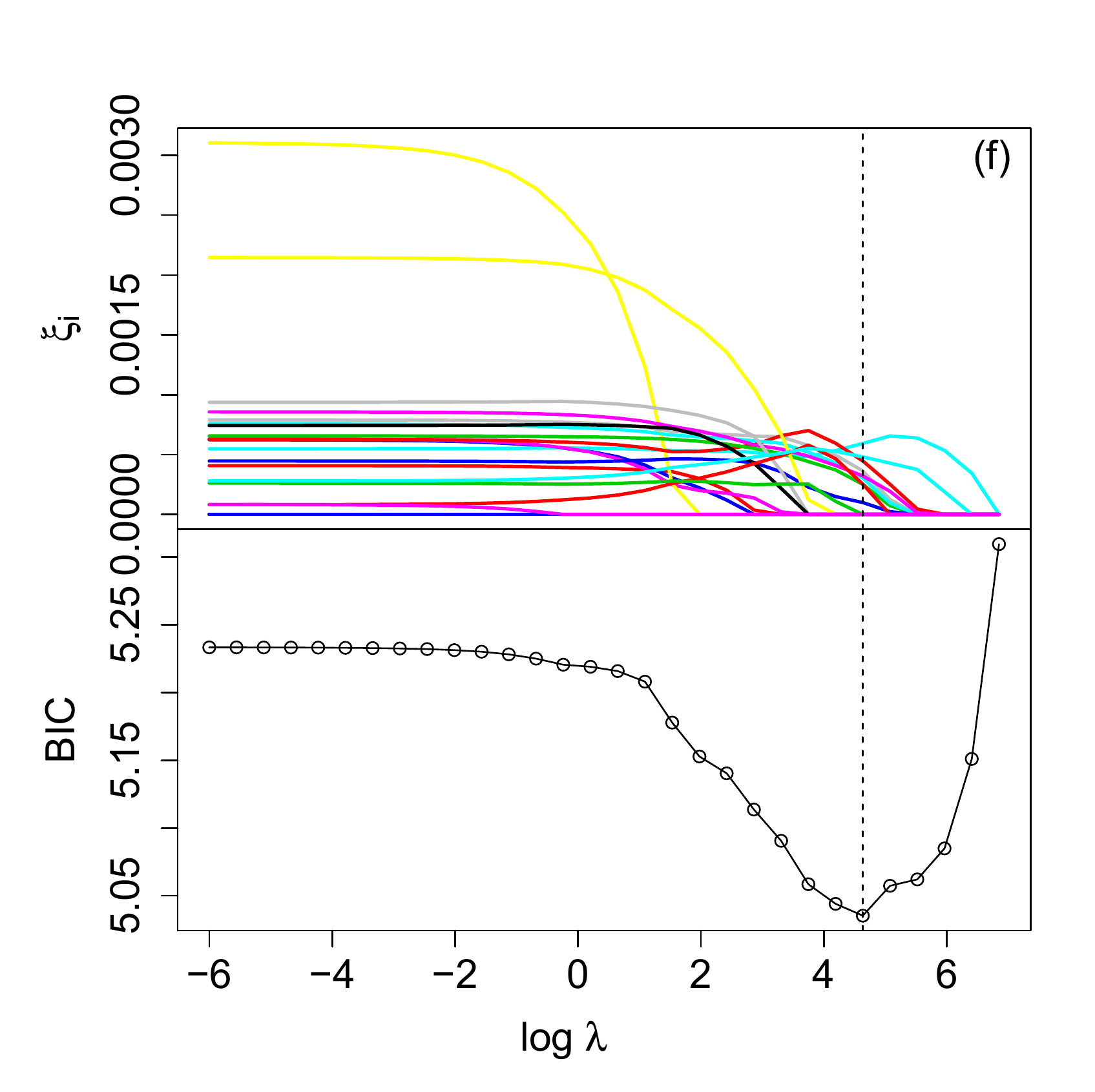}
\caption{NGK solution paths and BIC curves for diabetes data pathway 133, 4, and 140 using Gaussian kernel. Left side:  $\xi_j$'s vs. $L_1$ norm of $\xi_j$'s, Right side:  $\xi_j$'s and BIC vs. $\log\lambda$.}\label{f10}
\end{center}
\end{figure}

Figures \ref{f10}(a), (c) and (e) plot the solution paths of the $\xi_j$'s corresponding to genes for three pathways. Figures \ref{f10}(b), (d) and (f) show the BIC curves to select genes where a total of 13, 7 and 9 genes are selected, respectively. The index sets for the selected genes of the three pathways by the Gauss kernel NGK method are $\hat{\ma}_{133}=\{1, 4, 5, 14, 19, 23, 29, 31, 34, 41, 51, 53, 57\}$, $\hat{\ma}_4=\{8, 10, 11, 12, 13, 14, 21\}$ and $\hat{\ma}_{140}=\{5,  8, 10, 11, 12, 13, 14, 18, 21\}$. However, as discussed for the SCA experiment (Section \ref{sec6.1}) and simulation Example 3 (Section \ref{sec5.4}), variable selection depending on  single draw  may not be powerful even if the observation number is large. In the diabetes data, there are only 35 observations. So we need additional steps to increase selection power. In this section we propose using  a residual permutation procedure to repeat the variable selection process and counting the total frequency/probability of each predictor.
\begin{itemize}
\item \emph{Step 1}: Apply the Gaussian kernel NGK variable selection method to the original dataset using the backfitting algorithm introduced in Section \ref{sec3}, obtain the selected variables $\hat{\bxi}=(\hat{\xi}_j)^T_{j\in\hat{\mathcal{A}}}$. Use $\hat{\bxi}$ to fit the Gaussian kernel machine again to obtain new $\hat{\bal}$ and new ${\lambda}_0$ by REML such that $\hat{\mby}=K(\hat{\bxi})\hat{\bal}$. Obtain the residual $\hat{\bel}=\mby-\hat{\mby}$. Center $\hat{\bel}$ by subtracting its mean.
\item \emph{Step 2}: Permute the residual $\hat{\boldsymbol\epsilon}$ to get new $\hat{\boldsymbol\epsilon}^*$ and simulate outcomes as $\mathbf y^*=K(\hat{\bxi})\hat{\bal}+\hat{\bel}^*$.
\item \emph{Step 3}: Based on the new dataset $(\mathbf y^*, X)$ with fixed initial $\hat{\bal}$ and fixed ${\lambda_0}$, apply the NGK variable selection method    again and obtain the selected gene set.
\item \emph{Step 4}: Repeat Steps 2-3 for a large number of iterations (e.g. 3000 times).
\item \emph{Step 5}: Obtain the empirical probability/frequency of selecting each variable.
\end{itemize}

\begin{figure}[bth]
\begin{center}
\includegraphics[width=80mm]{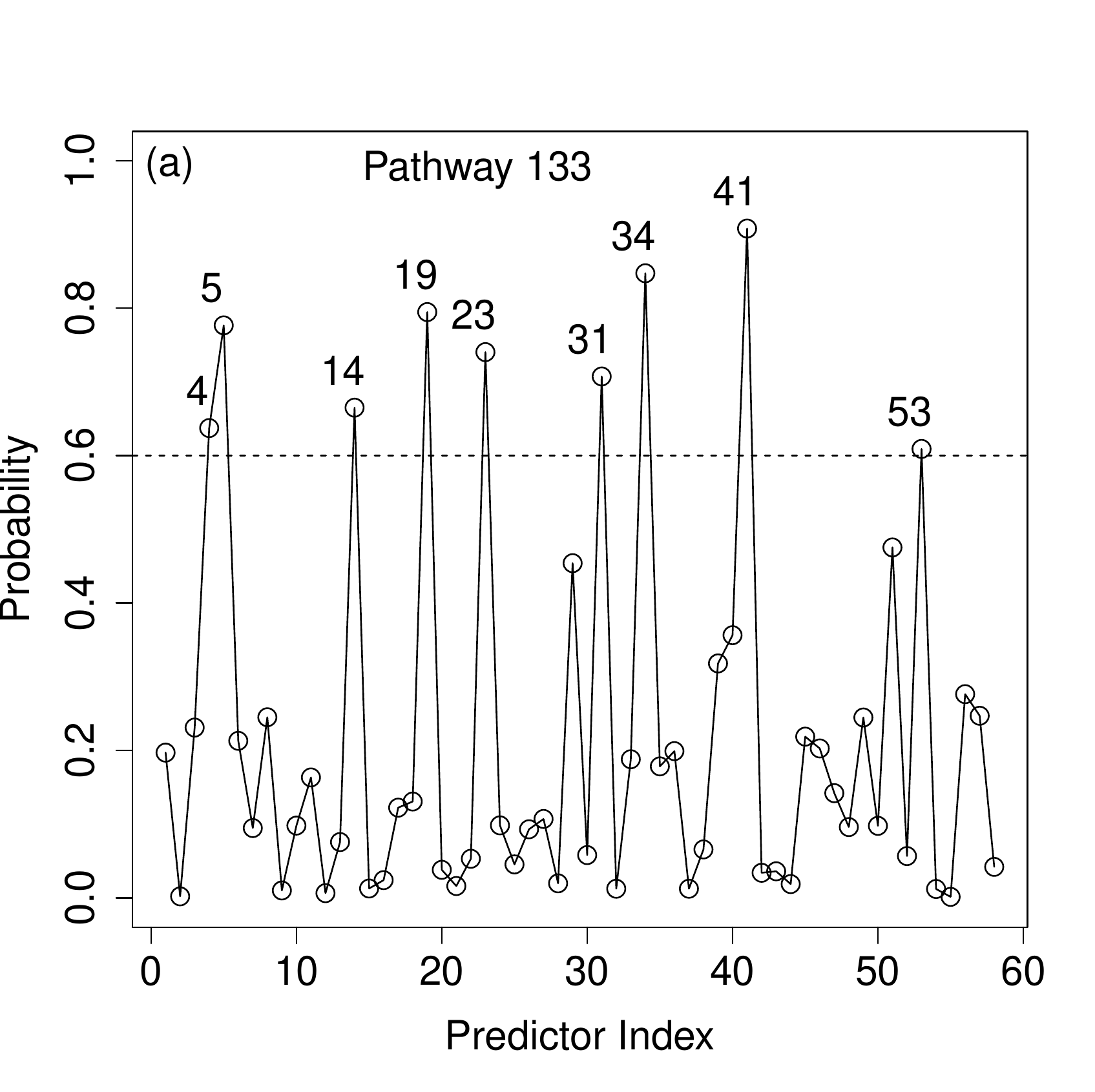}\includegraphics[width=80mm]{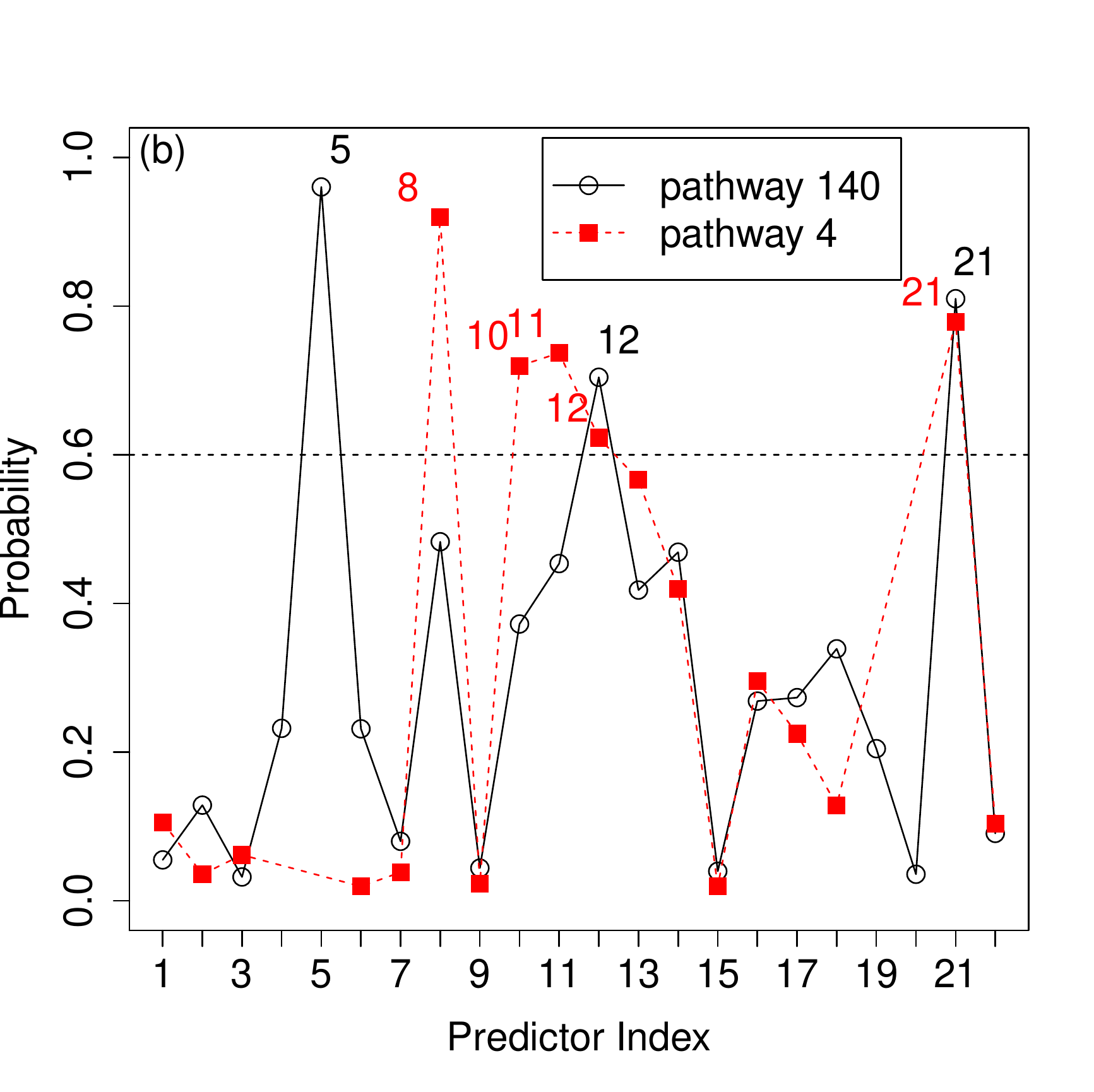}
\caption{Selection probability of each gene using the residual permutation method for pathway 133 (a), and pathway 140 and 4 (b), with a total of 3000 runs for each pathway.}\label{f11}
\end{center}
\end{figure}

The results of NGK permutation procedure are summarized in Figure \ref{f11}. If we take $60\%$ as the threshold, the sets of genes selected are ${\ma}_{133}^*=\{4, 5, 14, 19, 23, 31, 34, 41, 53\}$, ${\ma}_{4}^*=\{8, 10, 11, 12, 21\}$ and ${\ma}_{140}^*=\{5, 12, 21\}$ respectively. Because pathway 4 is a subset of pathway 140, we plot the results of the two pathways in one plot, Figure \ref{f11}(b). Compare with $\hat{\ma}$, we see that $\ma^*\subset\hat{\ma}$. For example, for pathway 133 four extra genes selected using a single NGK step are $\{1, 29, 51, 57\}$. Especially for gene 1, the selection probability is less than $20\%$ by permutation approach. %Again there is no much theory work on variable selection with residuals permutation, some references can be found at Knight and Fu (2000) and Chatterjee and Lahiri (2011) for the residuals bootstrapping LASSO estimator.

Interesting observations for pathway 4 and pathway 140 are found in Figure \ref{f11}(b). First, some of the genes are not significantly related to the response, such as genes $\{1, 2, 3, 7, 9, 15, 22\}$. In both pathways, the selection probabilities remains small for those genes. Another observation is that some genes are significantly related to the response and retain a high selection probability in both pathways (gene 21, for example). Furthermore, some genes seem to interact with one another. For example, genes $\{10, 11, 12, 13, 14\}$ appear to group a gene segment with similar selection probability. An interesting gene is gene 5, which does not appear in pathway 4. Gene 5 has the highest selection probability in pathway 140. While gene 5 is present in pathway 140, the selection probabilities of $\{8, 10, 11\}$ are smaller than in pathway 4. This indicates some interaction may occur between gene 5, gene 8 and the gene segment$\{10, 11\}$.

\section{Discussion}\label{sec7}
In this paper we have proposed a new variable selection approach to recover sparsity of the multivariate input variable in a nonadditive smoothing function. Our approach can be addressed as a nonnegative garrote variable selection procedure with kernel machine. The method we proposed has several advantages: (1) it can recover sparsity as well as model any order interactions automatically; (2) it is applicable not only to nonadditive smoothing functions, but also to additive model by choosing  a different kernel; and (3) it establishes a connection among several existing methods including linear nonnegative garrote, kernel learning and ARD problems.  We have also developed an efficient coordinate descent updating procedure for the scale parameters $\xi_j$'s which inherits the nice properties of the regular backfitting method and can replaces the LARS algorithm in models with additive multiple kernels.

The results in this paper show some theoretical properties similar to linear LASSO and linear nonnegative garrotes. However, other theoretical properties require further study, such as the convergence rate of the coordinate descent algorithm and the performance of model selection criteria such as BIC in least squares error kernel machines. Furthermore, in this paper, we suggested resampling variable selection procedures in two cases: when $n$ is large and when $n$ is small. Thus consistency and convergence rate of resampling/bootstrapping on NGK approaches are interesting future topics as well.

Possible extensions of our method include applying NGK approaches to generalized linear models (GLM). Logistic kernel machine regression or multiple categorical classification are popular for many applications. Selecting input variables using NGK applied to GLMs is challenging work as the link functions are nonlinear too.

Another interesting extension of our method is consideration of more complicated kernel structures. To illustrate this, we could consider a dataset with $q$ multivariate variables such as $q$ genetic pathways, each one containing multidimensional genetic expressions and potential genes sharing between pathways. Thus the kernel could be expressed as $K=\rho_1K_1(\bxi_1,X_1)+\rho_2K_2(\bxi_2,X_2)+...,\rho_qK_q(\bxi_q,X_q)$. Applying penalty on $\boldsymbol\rho=(\rho_1,...,\rho_q)^T$ and on $\bxi_1,...,\bxi_q$, NGK may be able to recover sparsity of the $X_j$'s as well as the additive functional components of $\{f_j=K_j\bal$\}'s. This might be considered as group NGK and we can apply it to selecting pathways and interactions from a pathway pool.

\Appendix
\numberwithin{equation}{subsection}
\section{}
\subsection{Proof of Lemma 2}\label{apen1}
{\it Proof:} This is a result of the composition theorem (Boyd and Vandenberghe 2004 Chp.3). For function $Q_0(\bxi)=Q_0\left(K(\bxi)\right)$ with domain $\textbf{dom }Q_0(\bxi)=\{\bxi \in \textbf{dom }K(\bxi)| K(\bxi)\in \textbf{dom }Q_0(K)\}$, if $Q_0(K)$ is convex and non-increasing and $K(\bxi)$ is concave, then $Q_0(\bxi)$ is convex. To see this, assuming $\bxi, \bxi'\in \textbf{dom }Q_0(\bxi)$, we have $\bxi, \bxi'\in \textbf{dom }K(\bxi)$ and $K(\bxi), K(\bxi')\in\textbf{dom } Q_0(K)$. Since $\textbf{dom } K(\bxi)=\mathbb{R}^p_+$ is convex, we have $\theta\bxi+(1-\theta)\bxi'\in\textbf{dom } K(\bxi)$ and, from concavity of $K(\bxi)$, we have
\be\label{e52}
K(\theta\bxi+(1-\theta)\bxi')\succeq\theta K(\bxi)+(1-\theta)K(\bxi').
\ee
Since $K(\bxi), K(\bxi')\in\mathbb{K}^*\subset\textbf{dom } Q_0(K)=\mathbb{K}$, we conclude that $\theta K(\bxi)+(1-\theta)K(\bxi')\in\textbf{dom } Q_0(K)$. Since $\theta\bxi+(1-\theta)\bxi'\in\textbf{dom } K(\bxi)$, we have $K(\theta\bxi+(1-\theta)\bxi')\in \textbf{dom } Q_0(K)$ too, which means $\theta\bxi+(1-\theta)\bxi'\in  \textbf{dom } Q_0(\bxi)$. Now, using the fact $Q_0(K)$ is nonincreasing and (\ref{e52}), we have
\be\label{e53}
Q_0\left(K(\theta\bxi+(1-\theta)\bxi')\right)\le Q_0\left(\theta K(\bxi)+(1-\theta)K(\bxi')\right).
\ee
Because of the convexity of $Q_0(K)$, we have
\be\label{e54}
Q_0\left(\theta K(\bxi)+(1-\theta)K(\bxi')\right)\le \theta Q_0\left(K(\bxi)\right)+(1-\theta)Q_0\left(K(\bxi')\right).
\ee
Combining the above two inequations, we get
\be\label{e55}
Q_0\left(\theta\bxi+(1-\theta)\bxi'\right)\le \theta Q_0\left(\bxi\right)+(1-\theta)Q_0\left(\bxi'\right),
\ee
which proves the convexity of $Q_0(\bxi)$. Since $\|\bxi\|_1$ is a convex function of $\bxi$, this implies convexity of $Q(\bxi)$.

\subsection{Proof of Theorem 1}\label{apen2}
{\it Proof:} We use the expression (\ref{e16}) to prove Theorem 1. Following Fan and Li (2001), to show the existence of a $d_n=n^{-1/2}$-consistent local minimum in the ball ${\bxi^*+d_n\mathbf u: \|\mathbf u\|\le C}$, we need to show that for any given $\epsilon>0$, there exists a large enough constant $C$ such that
\be\label{e56}
\lim\inf_{n}P\left\{\inf_{\|\mathbf u\|=C}Q(\bxi^*+d_n\mathbf u)>Q(\bxi^*)\right\}\ge1-\epsilon.
\ee
To show that, we first calculat following expression:
\be\label{e57}
\begin{split}
&Q(\bxi^*+d_n\mathbf u)-Q(\bxi^*)\\
&\approx d_n\left({{\partial Q_0}\over{\partial\bxi}}\right)_{\bxi^*}^T\mathbf u+{d^2_n\over2}\mathbf u^T\left({{\partial^2 Q_0}\over{\partial\bxi^T\partial\bxi}}\right)_{\bxi^*}\mathbf u+nd_n\lambda(\|\bxi^*+d_n\mathbf u\|-\|\bxi^*\|)\\
&\ge\sqrt{n}\lambda_0d_n\mathbf v_n(\bxi^*)^T\mathbf u+{{n\lambda_0d_n^2}\over2}\mathbf u^TM_n(\bxi^*)\mathbf u+nd_n\lambda\sum_{i=1}^{a}\left(|\xi_i+d_nu_i|-|u_i|\right)\\
&\ge\lambda_0\mathbf v_n^T(\bxi^*)\mathbf u+{\lambda_0\over2}\mathbf u^TM_n(\bxi^*)\mathbf u-\sqrt{n}\lambda a\|\mathbf u\|.
\end{split}
\ee
Using the regularity conditions of (\ref{e26}), we note that ${\mathbf v}_n^T=O_p(1)$. Thus in the right hand side of (\ref{e57}), the first term is uniformly bounded by second term for $C$ sufficiently large. To see this, at $\|\mathbf u\|=C$, $0.5{\mathbf u}^TM_n\mathbf u$ is uniformly larger than $0.5\Lambda_{min}(M_n)C^2$ which is a quadratic function of $C$ because $M_n$ is finite positive (semi-)definite. And $\|{\mathbf v}^T_n\mathbf u\|\le\|{\mathbf v}^T_n\|C$ which is linear function of $C$ since $\|{\mathbf v}_n^T\|=O_p(1)$. For sufficiently large $C$, the quadratic form of $C$ always dominates the linear form of $C$. As $n\rightarrow\infty$, we assume $\sqrt{n}\lambda\rightarrow0$, thus the last term is also bounded by the second term. Hence by choosing a sufficiently large $C$, (\ref{e56}) holds.

\subsection{Proof of Lemma 3}\label{apen3}
{\it Proof:} To continue the proof, (\ref{e39}) is a convex function of $\bxi$ by the Karush-Kuhn-Tucker conditions for optimality in a convex program, the point $\hat{\bxi}\in\mathbb{R}^p_+$ is optimal if and only if there exists a subgradient $\hat{\bka}\in \partial(\|\hat{\bxi}\|_1)$ such that
\be\label{e58}
\left.{{\partial\tilde Q_0}\over{\partial\bxi}}\right|_{\hat\bxi}+n\lambda\hat{\bka}=0,
\ee
where $\tilde Q_0$ is the first two terms of (\ref{e38}). The collection of subgradient of $\|\hat{\bxi}\|_1$ at point $\hat{\bxi}$ is the subdifferential $\partial(\|\hat{\bxi}\|_1)$:
\be\label{e59}
\partial(\|\hat{\bxi}\|_1)=\{\hat{\bka}\in\mathbb{R}^p: \hat\kappa_j=1 \textrm{ for } \hat\xi_j>0; \hat\kappa_j\le1 \textrm{ for } \hat\xi_j=0\}.
\ee
Plugging back into (\ref{e39}) and operating simple algebra, we have
\be\label{e60}
Z^TZ(\hat{\bxi}-\bxi^*)-Z^T\bel+{\lambda_0\over2}Z^T\bal^*+n\lambda\hat{\bka}=0.
\ee
Suppose $\lim_nP(\hat{\ma}=\ma)\rightarrow1$ or $\lim_nP(\sgn(\hat\bxi)=\sgn(\bxi^*))\rightarrow1$, thus
\[\label{e61}
\hat{\bxi}_1\succ\mathbf0, \hat{\bxi}_0=\mathbf0,\; \hbox{and}\; \hat{\bka}_1=\mathbf 1, \hat{\bka}_0\preceq\mathbf1.
\]
Substituting these observations and rearranging (\ref{e39}), we have
\begin{subequations}
\be\label{e62a}
{1\over n}Z^T_0\left(\bel-{\lambda_0\over2}\bal^*\right)-Z^T_0Z_1(Z^T_1Z_1)^{-1}\left[{1\over n}Z^T_1\left(\bel-{\lambda_0\over2}\bal^*\right)-\lambda\mathbf 1\right]=\lambda\hat{\bka}_0,
\ee
\be\label{e62b}
\bxi^*_1+\left({1\over n}Z^T_1Z_1\right)^{-1}\left[{1\over n}Z^T_1\left(\bel-{\lambda_0\over2}\bal^*\right)-\lambda\mathbf 1\right]=\hat{\bxi}_1.
\ee
\end{subequations}
Considering conditions (\ref{e59}) for $\hat\bka_0$ and $\hat{\bxi}_1\succ\mathbf0$, we have the sufficient and necessary conditions of (\ref{e37a}) and (\ref{e37b}).

\subsection{Proof of Theorem 2}\label{apen4}
{\it Proof:} Condition $|\tilde{\bal}-\bal^*|_\infty=O_p(\delta)$ implies that $|\tilde\alpha_k\tilde\alpha_l-\alpha^*_k\alpha^*_l|\le(|\tilde\alpha_k|+|\alpha^*_l|)|\tilde\alpha_l-\alpha^*_l|=O_p(\delta)$ for $1\le k,l\le n$, thus we have the relationships
\be\label{e63}
\begin{split}
&n^{-1}\tilde{\mathbf z}^T_i\tilde{\mathbf z}_j=n^{-1}{\mathbf z}_i^T{\mathbf z}_j+O_p(\delta),\\
&n^{-1}\tilde{\mathbf z}^T_j\tilde\bal=n^{-1}{\mathbf z}_j^T\bal^*+O_p(\delta),
\end{split}
\ee
where $1\le i, j\le p$. These relationships are derived from the following two inequalities:
\be\label{e64}
\begin{split}
\left|{1\over n}\left\{\tilde {\mathbf z}_i^T\tilde{\mathbf z}_j-{\mathbf z}_i^T{\mathbf z}_j\right\}\right|
&=\left|{1\over n}\left\{\tilde\bal^TK'_iK'_j\tilde\bal-{\bal^*}^TK'_iK'_j\bal^*\right\}\right|\\
&=\left|{1\over n}\left\{\sum_{k,l}(K'_iK'_j)_{k,l}(\tilde\alpha_k\tilde\alpha_l-\alpha^*_k\alpha^*_l)\right\}\right|\\
&\le{1\over n}\left\{\sum_{k,l}|K'_iK'_j|_{k,l}|\tilde\alpha_k\tilde\alpha_l-\alpha^*_k\alpha^*_l|\right\}\\
&\le{{O_p(\delta)}\over n}\sum_{k,l}|K'_iK'_j|_{k,l}= O_p(\delta){{\mathbf 1^T|K'_iK'_j|\mathbf 1}\over n}\\
&\le O_p(\delta)\cdot C=O_p(\delta)
\end{split}
\ee
and
\be\label{e65}
\begin{split}
\left|{1\over n}\left\{\tilde{\mathbf z}_j^T\tilde\bal-{\mathbf z}_j^T\bal^*\right\}\right|
&=\left|{1\over n}\left\{\tilde\bal^TK'_j\tilde\bal-{\bal^*}^TK'_j\bal^*\right\}\right|\\
&=\left|{1\over n}\left\{\sum_{k,l}(K'_j)_{k,l}(\tilde\alpha_k\tilde\alpha_l-\alpha^*_k\alpha^*_l)\right\}\right|\\
&\le{1\over n}\left\{\sum_{k,l}|K'_j|_{k,l}|\tilde\alpha_k\tilde\alpha_l-\alpha^*_k\alpha^*_l|\right\}\\
&\le O_p(\delta){{\mathbf 1^T |K'_j|\mathbf 1}\over n}\\
&\le O_p(\delta)\cdot C=O_p(\delta),
\end{split}
\ee
where the $C$'s are some small positive numbers. These inequalities are true for Gaussian and linear polynomial kernels because $X$ is standardized.  For example, for a Gaussian kernel, we have $n^{-1}\mathbf 1^T|K'_j|\mathbf 1\le 2$ where $|\cdot|$ is the componentwise absolute value. To see this, first note that $\mathbf 1^T|K_j'|\mathbf 1=\mathbf 1^T|K\circ D^j|\mathbf 1\le\mathbf1^T|D^j|\mathbf1$ since all elements of $K$ are positive and smaller than 1, and $\mathbf1^T|D^j|\mathbf1=\sum_{k,l}(x_{jk}-x_{jl})^2=\sum_{k,l}(x^2_{jk}+x^2_{jl}-2x_{jk}x_{jl})=\sum_k(nx^2_{jk}+1)=2n$. For a linear polynomial kernel, $K'_j=D^j=\mbx_j\mbx_j^T$. Thus $\mathbf 1^T|K_j'|\mathbf 1=\sum_{k,l}|x_{jk}x_{jl}|\le(\sum_l|x_{jl}|)^2\le n\sum_lx_{jl}^2=n$. In both cases we use $\sum_lx_{jl}=0$ and $\sum_lx^2_{jl}=1$. Similarly we can show that the inequalities for $n^{-1}\mathbf 1^T|K'_iK'_j|\mathbf 1$ are bounded by some small numbers.

In addition, from conditions (\ref{e41a})-(\ref{e41c})  and the relationships (\ref{e63}), with $\delta\rightarrow0$, the left hand side of (\ref{e41a}) only differs from $\tilde\Sigma_{01}\tilde\Sigma_{11}^{-1}\mathbf 1-{\lambda_0\over{2n\lambda}}\tilde Z_0^T\tilde P\tilde\bal$ by $O_p(\delta)$ ($\tilde P$ is the projection matrix and thus does not change much the norm). For $\delta$ sufficiently small,
\be\label{e66}
\tilde\Sigma_{01}\tilde\Sigma_{11}^{-1}\mathbf 1-{\lambda_0\over{2n\lambda}}\tilde Z_0^T\tilde P\tilde{\bal}\preceq(1-\tilde\gamma)\mathbf 1
\ee
holds for some $\tilde\gamma\in(0,1]$. The converse is also true: if \ref{e66} satisfied, then we can always find a small positive $\gamma$ when $\delta\rightarrow0$ such that the condition (\ref{e41a}) is true. This equivalence allows us to show sparsistency by using (\ref{e66})

Starting from (\ref{e66}), our argument is based on the technique of a primal dual witness on model selection, consistency of the lasso which contains the following steps (Wainwright, 2009):
\begin{enumerate}
\item Obtain $\hat{\bxi}_1$ by solving (\ref{e40b}), and set $\hat{\bxi}_0=0$,
\item Set $\hat{\bka}_1=\partial(\|\bxi^*_1\|_1)$, for our model with nonnegative garrotte $\hat{\bka}_1=\mathbf 1$,
\item With these setting of $\hat{\bxi}_1$ and $\hat{\bka}_1$, obtain $\hat{\bka}_0$ through (\ref{e40a}), and check whether or not $\hat\bka_0\in\partial(\|\bxi^*_0\|_1)$, for for our model with nonnegative garrotte $\hat\bka_0\prec\mathbf1$,
\item Check whether $\hat{\bka}_1=\mathbf 1$.
\end{enumerate}
Lemma 2 in Wainwright (2009) states that if dual feasibility is established (Step 1-3 succeed), then $\hat{\ma}\subseteq\ma$. In Step 3 using $\hat\bka_0\prec\mathbf1$ instead of $\hat\bka_0\preceq\mathbf1$ ensures uniqueness by strict dual feasibility. Furthermore, if Step 4 succeeds as well, then $\hat{\ma}=\ma$.

Following Wainwright (2009), Theorem 2 is proved in two steps.  Given $\tilde{\bal}$ and $\tilde Z$ defined as before, from (\ref{e40a}-\ref{e40b}), we define two random variables:
\begin{align*}\label{e67}
A_i&:=\tilde {\mathbf z}_i^T\left\{\tilde Z_1(\tilde Z_1^T\tilde Z_1)^{-1}\mathbf 1-{\lambda_0\over{2n\lambda}}\tilde P\tilde{\bal}\right\}+{1\over{n\lambda}}\tilde {\mathbf z}_i^T\tilde P\bel,\;i\in\ma^c\\
\hat\xi_j-\xi^*_j&:={\mathbf e}^T_j\tilde\Sigma_{11}^{-1}\left({1\over{n}}\tilde  Z_1^T\bel\right)-{\mathbf e}^T_j\tilde\Sigma_{11}^{-1}\left\{{\lambda_0\over{2n}}\tilde Z_1^T\tilde{\bal}+\lambda\mathbf 1\right\},\;j\in\ma,
\end{align*}
where ${\mathbf e}_j^T$ is the selection vector with 1 in the $j$th position.
\begin{itemize}
\item {\it Dual feasibility}\\
Write $A_i$ as $E(A_i)+A_i^*$, where $E(A_i)=\tilde {\mathbf z}_i^T\left\{\tilde Z_1(\tilde Z_1^T\tilde Z_1)^{-1}\mathbf 1-{\lambda_0\over{2n\lambda}}\tilde P\tilde{\bal}\right\}$, and $A_i^*={1\over{n\lambda}}\tilde {\mathbf z}_i^T\tilde P\bel$. To have the subgradient vector $\hat{\bka}_0\preceq\mathbf 1$ is equivalent to showing
\[\label{e68}
\max_iA_i\le1.
\]
Using the definition of $A_i$ and condition (\ref{e66}), we have
\[\label{e69}
\max_iA_i\le(1-\tilde\gamma)+\max_iA_i^*.
\]
$A_i^*$ is a zero mean sub-Gaussian random variable and, according to Wainwright (2009), the variance of $A_i^*$ is bounded by
\[\label{e70}
\begin{split}
\var{(A_i^*)}={\sigma^2\over{\lambda^2n^2}}(\tilde {\mathbf z}_i^TP\tilde {\mathbf z}_i)&\le{\sigma^2\over{\lambda^2n^2}}\|\tilde {\mathbf z}_i\|_2^2={\sigma^2\over{\lambda^2n}}\left(n^{-1}\|{\mathbf z}_i\|_2^2+O_p(\delta)\right)\\
&\le{\sigma^2\over{\lambda^2n}}(1+O_p(\delta)),
\end{split}
\]
which can be shown using the relationship (\ref{e63}), the properties of the projection matrix, and normalized ${\mathbf z}_i$ vector such that $\|{\mathbf z}_i\|^2_2\le n$, and $\delta\rightarrow0$.

By the sub-Gaussian tail bound results combined with the union bound (Wainwright 2009), we have
\[\label{e71}
\begin{split}
P\left(\max_iA_i^*\ge{\tilde\gamma\over2}\right)&\le(p-a)\exp\left\{-{{(\tilde\gamma/2)^2}\over{2\sigma^2\lambda^{-2}n^{-1}[1+O_p(\delta)]}}\right\}\\
&=\exp\left\{-{{\lambda^2n{\tilde\gamma}^2}\over{8\sigma^2}}(1+O_p(\delta))^{-1}+\log(p-a)\right\}.
\end{split}
\]
Putting all these parts together, we conclude that
\[\label{e72}
P\left(\max_iA_i>1-{\tilde\gamma\over2}\right)\le\exp{\left(-\eta_1\lambda^2n\right)}.
\]
If we choose some $\lambda$ such that ${{\lambda^2n{\tilde\gamma}^2}\over{8\sigma^2}(1+O_p(\delta))}>\log(p-a)$, say
\be\label{e73}
\lambda>{2\over\tilde\gamma}\sqrt{{{2\sigma^2\log p}\over n}\left(1+O_p(\delta)\right)},
\ee
the probability for $\{\max_iA_i>1-\tilde\gamma/2\}$ vanishes with rate $\exp(-\eta_1\lambda^2n)$ as $n\rightarrow\infty$. Or in other words, with probability $1-\exp(-\eta_1\lambda^2n)$, we have $\hat{\ma}\subseteq\ma$.
\item {\it Bounding $\|\hat{\bxi}_1-\bxi^*_1\|_\infty$}\\
The upper bound of $\left\|\hat{\bxi}_1-\bxi^*_1\right\|_\infty$ is
\bea\label{e74}
\|\hat{\bxi}_1-\bxi^*_1\|_\infty\le\underbrace{\left\|\tilde\Sigma_{11}^{-1}\left({1\over n}\tilde Z_1^T\bel\right)\right\|_\infty}_I+\underbrace{\left\|\tilde\Sigma_{11}^{-1}\left({{\lambda_0}\over {2n}}\tilde Z_1^T\tilde{\bal}\right)\right\|_\infty}_{II}+\underbrace{\lambda\|\tilde\Sigma_{11}^{-1}\|_\infty}_{III}.
\eea
Note the $\infty-$norm of matrix $\tilde\Sigma_{11}^{-1}$ is bounded as
\be\label{e75}
\|\tilde\Sigma_{11}^{-1}\|_\infty\le\sqrt{a}\tilde C_{min}^{-1}.
\ee
Thus, part III is bounded as $\lambda\|\tilde\Sigma_{11}^{-1}\|_\infty=\sqrt{a}\lambda\tilde C_{min}^{-1}$.

Part II can be bounded as
\be\label{e77}
\begin{split}
II:=&\left\|\tilde\Sigma_{11}^{-1}\left({{\lambda_0}\over {2n}}\tilde Z_1^T\tilde{\bal}\right)\right\|_\infty\le\|\tilde\Sigma_{11}^{-1}\|_\infty\left\|{\lambda_0\over{2n}}\tilde Z_1^T\tilde{\bal}\right\|_\infty\\
=&\|\tilde\Sigma_{11}^{-1}\|_\infty\left\|{\lambda_0\over\sqrt{n}}\left({1\over{2\sqrt{n}}}Z_1^T\bal^*+O_p(\delta\sqrt{n})\right)\right\|_\infty\\
\le&\|\tilde\Sigma_{11}^{-1}\|_\infty\cdot{\lambda_0}\left(n^{-1/2}\max_{j}|v^*_j|+O_p(\delta)\right),
\end{split}
\ee
where we use (\ref{e63}) and (\ref{e27}) for $\mathbf v^*$. Using (\ref{e75}) we have
\be\label{e78}
II\le{{\sqrt{a}\lambda_0}\over\tilde C_{min}}\left(n^{-1/2}\max_j|v^*_{j}|+O_p(\delta)\right).
\ee
Note that in $\hat\xi_j-\xi^*_j$, the random portion is $U_j:={\mathbf e}_j^T\tilde\Sigma_{11}^{-1}(n^{-1}\tilde Z_1^T\bel)$ with $\bel\sim N(0, \sigma^2I)$, so $U_j$ is zero mean Gaussian, i.e. $E(U_j)=0$, and
\be\label{e79}
\var{(U_j)}={\sigma^2\over n}{\mathbf e}_j^T\tilde\Sigma_{11}^{-1}{\mathbf e}_j\le{\sigma^2\over n}\tilde C_{min}^{-1}.
\ee
Again using the sub-Gaussian tail bound (Wainwright 2009), we have
\be\label{e80}
\begin{split}
P\left(\max_j|U_j|>t\right)&\le2\exp\left(-{{t^2n}\over{2\sigma^2\tilde C_{min}^{-1}}}+\log{a}\right)\\
&=2\exp\left(-{{t^2n}\over{2\sigma^2}}\tilde C_{min}+\log{a}\right).
\end{split}
\ee
Setting $t=4\sigma\lambda \tilde C_{min}^{-1/2}$, and by choosing $\lambda$ as in (\ref{e73}), we have $8\lambda^2n>\log p\ge\log a$ so that $P\left(\max_j|U_j|>4\sigma\lambda \tilde C_{min}^{-1/2}\right)\rightarrow0$ with rate at least $2\exp(-\eta_2\lambda^2n)$ where $\eta_2>0$. And we are bounding
\[\label{e81}
\begin{split}
\|\hat{\bxi}_1-\bxi^*_1\|_\infty&\le\lambda\left[{{4\sigma}\over \sqrt{\tilde C_{min}}}+\|\tilde\Sigma_{11}^{-1}\|_\infty\cdot{\lambda_0\over{\lambda}}\left(n^{-1/2}\max_{j}|v^*_j|+O_p(\delta)\right)+\|\tilde\Sigma^{-1}_{11}\|_\infty\right]\\
&\le\lambda\left[{{4\sigma}\over \sqrt{\tilde C_{min}}}+{{\lambda_0\sqrt{a}}\over{\lambda \tilde C_{min}}}\left(n^{-1/2}\max_j|v^*_{j}|+O_p(\delta)\right)+{\sqrt{a}\over \tilde C_{min}}\right]
\end{split}
\]
with probability $1-2\exp(-\eta_2\lambda^2n)$.\\
From the bounding expression, we can see that if we have $\lambda+{\lambda_0\sqrt{a/n}}+O_p(\lambda_0\sqrt{a}\delta)+\sqrt{a}\lambda\rightarrow0$ and $\lambda^2n\rightarrow\infty$, the we have $\hat\bxi_1\rightarrow\bxi^*_1$ with probability 1.\\
Furthermore define
\[\label{e82}
\rho(\lambda)=\lambda\left[{{4\sigma}\over \sqrt{\tilde C_{min}}}+\|\tilde\Sigma_{11}^{-1}\|_\infty\cdot{\lambda_0\over{\lambda}}\left(n^{-1/2}\|\mathbf v^*\|_\infty+O_p(\delta)\right)+\|\tilde\Sigma^{-1}_{11}\|_\infty\right].
\]
Hence we finally conclude that, as $\lambda^2n\rightarrow\infty$, if $\rho(\lambda)<\min_{j\in\ma}\xi^*_j$, then we have all $\hat\xi_j>0, j\in\ma$, thus establishing the sign consistency $\hat{\ma}=\ma$.
\end{itemize}

%%%%%%%%%%% referece starts here %%%%%%%%%%%%
\vskip 5mm
\noindent
{\large{\bf REFERENCES}}
%\begin{thebibliography}{}
%\bibitem{ }
\refmark Bach, F. (2008). Consistency of the Group Lasso and Multiple Kernel Learning. \JMLR, 9, 1179-1225.
\refmark Boyd, S. and Vandenberghe, L. (2004). {\it Convex Optimization}. New York: Cambridge University Press.
\refmark Breiman, L. (1995). Better Subset Regression Using the Nonnegative Garrote. {\it Technometrics}, 37, 373-384.
%\refmark Chatterjee, A. and Lahiri, S. N. (2011). Bootstrapping Lasso Estimators. \JASA, 106, 608-625.
\refmark Fan, J. and Li, R. (2001). Variable Selection via Nonconcave Penalized Likelihood and its Oracle Properties. \JASA, 96, 1348-1360.
\refmark Fan, J. and Lv, J. (2008). Sure Independence Screening for Ultrahigh Dimensional Feature Space. \JRSSB, 70, 849-911.
\refmark Fan, J., Feng, Y. and Song, R. (2011) Nonparametric Independence Screening in Sparse Ultra-High-Dimensional Additive Models. \JASA, 106, 544-557.
\refmark Green, P. J. and Silverman, B. W. (1994). {\it Nonparametric Regression and Generalized Linear Models}. London: Chapman and Hall.
\refmark Hall, P., Lee, E. R. and Park, B. U. (2009). Bootstrap-Based Penalty Choice for the Lasso, Achieving Oracle Performance. \SSS, 19, 449-471.
\refmark Hastie, T. and Tibshirani, R. (1990). {\it Generalized Additive Models}. London; New York: Chapman and Hall.
\refmark Krishnapuram, B., Hartemink, A. J. and Carin L. (2004). A Bayesian Approach to Joint Feature Selection and Classifier Design. {\it IEEE Transactions on Pattern Analysis and Machine Intelligence}, 26, 1105-1111.
\refmark Kimeldorf, G. and Wahba, G. (1971). Some Results on Tchebychefian Spline Functions. {\it Journal of Mathematical Analysis and Applications}, 33, 82-95.
%\refmark Knight, K. and Fu, W. (2000). Asymptotics for Lasso-Type Estimators. \ANNALS, 28, 1356-1378.
\refmark Lanckriet, G., Cristianini, N., Bartlett, P., Ghaoui, L. E. and Jordan, M. I. (2004). Learning the Kernel Matrix with Semi-Definite Programming. \JMLR, 5, 27-72
\refmark Lin, Y. and Zhang, H. H. (2006). Component Selection and Smoothing in Multivariate Nonparametric Regression. \ANNALS, 34, 2272-2297.
\refmark Linkletter, C., Bingham, D., Hengartner, N., Higdon, D. and Ye K. Q. (2006). Variable Selection for Gaussian Process Model in Computer Experiments. \TECH, 48, 478-490.
\refmark Liu, D., Lin, X. and Ghosh, D. (2007). Semiparametric Regression of Multi-Dimensional Genetic Pathway Data: Least Squares Kernel Machines and Linear Mixed Models.  \BMCS, 63, 1079-1088.
\refmark MacKay, D. J. C. (1994). Bayesian Methods for Backprop Networks. In Domany, E., van Hemmen,  J. L. and Schulten, K., editors, {\it Models of Neural Networks, III}, Chapter 6, 211-254. Springer.
%\refmark Meinshausen, N. and B\"{u}hlmann P. (2006). High-Dimensional Graphs and Variable Selection with the Lasso. \ANNALS, 34, 1436-1462.
\refmark Micchelli, C. A. and Pontil, M. (2005). Learning the Kernel Function via Regularization. \JMLR, 6, 1099-1125.
\refmark Mootha, V. K., Lindgren, C. M., Eriksson, K., Subramanian, A., Sihag, S., Lehar, J., Puigserver, P., Carlsson, E., Ridderstrale, M., Laurila, E., Houstis, N., Daly, M. J., Patterson, N., Mesirov, J. P., Golub, T. R., Tamayo, P., Spiegelman, B., Lander, E. S., Hirschhorn, J. N., Altshuler, D. and Groop, L. C. (2003). PGC-l alpha-Responsive Genes Involved in Oxidative Phosphorylation are Coordinately Downregulated in Human Diabetes. {\it Nature Genetics}, 34, 267-273.
\refmark Neal, R. M. (1996) {\it Bayesian Learning for Neural Networks, Lecture Notes in Statistics No. 118}, New York: Springer-Verlag.
\refmark Radchenko, P. and James, G. M. (2010). Variable Selection Using Adaptive Nonlinear Interaction Structures in High Dimensions. \JASA, 105, 1541-1553.
\refmark Rakotomamonjy, A., Bach, F., Canu, S. and Grandvalet, Y. (2008). SimpleMKL. \JMLR, 9, 2491-2521.
%\refmark Rasmussen, C. E. and Williams, C. K. I. (2006). {\it Gaussian Process for Machine Learning}. Cambridge: MIT Press.
\refmark Ravikumar, P., Lafferty, J., Liu, H. and Wasserman, L. (2009). Sparse Additive Models. \JRSSB, 71, 1009-1030.
%\refmark Tipping, M. E. (2001). Sparse Bayesian Learning and the Relevance Vector Machine. \JMLR, 1, 211-244.
\refmark Savitsky, T., Vannucci, M. and Sha N. (2011). Variable Selection for Nonparametric Gaussian Process Priors: Models and Computational Strategies. \STATSCI, 26, 130-149.
\refmark Wang, H. and Leng, C. (2007). Unified LASSO Estimation by Least Squares Approximation. \JASA, 102, 1039-1048.
\refmark Wahba, G. (1990).  {\it Spline Models for Observational Data}. Philadelphia: Society for Industrial and Applied Mathematics.
\refmark Wainwight, M. (2009). Sharp Thresholds for High-Dimensional and Noisy Sparsity Recovery Using $l_1$-Constrained Quadratic Programming (Lasso). {\it IEEE Transactions on Information Theory}, 55, 2183-2202.
\refmark Yuan, M. (2007). Nonnegative Garrote Component Selection in Functional ANOVA Models. {\it Proceedings of AI and Statistics, AISTATS}, 660-666.
\refmark Yuan, M. and Lin, Y. (2007). On the Nonnegative Garrote Estimator. \JRSSB, 69, 143-161.
\refmark Zhao, P. and Yu, B. (2006). on Model Selection Consistency of Lasso. \JMLR, 7, 2541-2563.
\refmark Zou, F., Huang, H., Lee, S. and Hoeschele, I. (2010). Nonparametric Bayesian Variable Selection with Applications to Multiple Quantitative Trait Loci Mapping with Epistasis and Gene-Environment Interaction. {\it Genetics}, 186, 385-394.
%\refmark Zou, H. (2006). The Adaptive Lasso and Its Oracle Properties. \JASA, 101, 1418-1429.

%%%%%%%%%%%%attached graphs and tables start here %%%%%%%%%%%%%%%%%%%

\pagebreak\clearpage\newpage
\thispagestyle{empty}

\begin{table}[h]
\centering
\caption{Simulation results of Simulation Example 1 for 200 runs. ``NGK Gauss'' and ``NGK Poly'' represent NGK method with Gaussian kernel and with linear polynomial kernel, respectively.}
\begin{tabular}{llccccc}
\hline\hline
                         &      & FP-rate    & FN-rate    &  MS          &  RSS       & SE        \\
\hline
\multirow{3}{*}{$n=64$}  &NGK Gauss & 0.12(0.11) & 0.20(0.15) & 4.46(1.77)   & 1.29(0.71) & 0.55(0.63) \\
                         &NGK Poly  & 0.08(0.10) & 0.20(0.12) & 4.26(1.15)   & 0.92(0.19) & 0.14(0.07) \\
                         &COSSO     & 0.09(0.10) & 0.19(0.12) & 4.42(1.24)   & 1.20(0.30) & 1.01(0.18) \\
\hline
\multirow{3}{*}{$n=128$} &NGK Gauss & 0.09(0.09) & 0.22(0.14) & 3.98(1.56)   & 1.13(0.36) & 0.26(0.33) \\
                         &NGK Poly  & 0.06(0.08) & 0.21(0.12) & 3.96(1.07)   & 0.96(0.12) & 0.07(0.04) \\
                         &COSSO     & 0.05(0.08) & 0.21(0.12) & 3.87(1.10)   & 1.04(0.15) & 1.00(0.12) \\
\hline
\multirow{2}{*}{$n=256$} &NGK Gauss & 0.07(0.09) & 0.22(0.15) & 3.83(1.65)   & 1.10(0.31) & 0.16(0.29)  \\
                         &NGK Poly  & 0.05(0.08) & 0.24(0.10) & 3.68(0.98)   & 0.96(0.08) & 0.04(0.02) \\
                         &COSSO     & 0.04(0.07) & 0.18(0.13) & 4.03(1.07)   & 1.00(0.10) & 1.00(0.09) \\
                         \hline
\hline
\end{tabular}
\label{t2}
\end{table}

\pagebreak\clearpage\newpage
\thispagestyle{empty}

\begin{table}[h]
\centering
\caption{Simulation results of Simulation Example 2 for 200 runs.}
\begin{tabular}{llccccc}
\hline\hline
                         &            & FP-rate    & FN-rate    &  MS          &  RSS       & SE        \\
\hline
\multirow{3}{*}{$n=64$}  &NGK Gauss & 0.09(0.11) & 0.00(0.00) & 5.59(0.83)   & 1.02(0.24) & 0.34(0.09) \\
                         &NGK Poly    & 0.05(0.09) & 0.00(0.00) & 5.34(0.56)   & 1.14(0.20) & 0.35(0.08) \\
                         &COSSO       & 0.04(0.08) & 0.00(0.00) & 5.32(0.61)   & 0.84(0.20) & 0.99(0.18) \\
\hline
\multirow{3}{*}{$n=128$} &NGK Gauss & 0.02(0.05) & 0.00(0.00) & 5.01(0.31)   & 1.15(0.17) & 0.27(0.05) \\
                         &NGK Poly    & 0.04(0.07) & 0.00(0.00) & 5.23(0.46)   & 1.20(0.15) & 0.31(0.05)  \\
                         &COSSO       & 0.01(0.04) & 0.00(0.00) & 5.06(0.27)   & 0.95(0.14) & 1.02(0.13) \\
\hline
\multirow{3}{*}{$n=256$} &NGK Gauss & 0.01(0.03) & 0.00(0.00) & 5.04(0.18)   & 1.12(0.12) & 0.20(0.05) \\
                         &NGK Poly    & 0.01(0.03) & 0.00(0.00) & 5.03(0.17)   & 1.22(0.11) & 0.29(0.03)  \\
                         &COSSO       & 0.01(0.03) & 0.00(0.00) & 5.05(0.21)   & 0.98(0.09) & 1.01(0.09) \\
\hline
\hline
\end{tabular}
\label{t4}
\end{table}

\pagebreak\clearpage\newpage
\thispagestyle{empty}

\begin{table}[h]
\centering
\caption{Simulation results of Simulation Example 3 for 400 runs.}
\begin{tabular}{llccccc}
\hline\hline
                         &            & FP-rate    & FN-rate    &  MS          &  RSS       & SE        \\
\hline
\multirow{2}{*}{$n=64$}  &NGK Gauss & 0.08(0.04) & 0.003(0.022) & 11.88(4.21)   & 1.56(0.30) & 1.01(0.20) \\
                         &NGK Poly    & 0.08(0.11) & 0.030(0.110) & 12.52(14.5)   & 1.27(1.53) & 0.87(1.37) \\
\hline
\hline
\end{tabular}
\label{t5}
\end{table}

\pagebreak\clearpage\newpage
\thispagestyle{empty}

\pagebreak\clearpage\newpage
\thispagestyle{empty}

\pagebreak\clearpage\newpage
\thispagestyle{empty}

\pagebreak\clearpage\newpage
\thispagestyle{empty}

\pagebreak\clearpage\newpage
\thispagestyle{empty}

\pagebreak\clearpage\newpage
\thispagestyle{empty}

\label{lastpage}

%%%%%%%%%%%%%%%%%%%%%%%%%%%%%%%%%%%%%%%%%%%%%%%%%%%%%%%%%%%%%%%%%%%%%%%%%
\end{document}